\documentclass{aa}
\usepackage[varg]{txfonts}
\usepackage{graphicx}
\usepackage{natbib,twoopt}
\usepackage[breaklinks=true]{hyperref} %% to avoid \citeads line fills
\usepackage{lscape}
\bibpunct{(}{)}{;}{a}{}{,} %% natbib format for A&A and ApJ
\makeatletter
 \newcommandtwoopt{\citeads}[3][][]{\href{https://ui.adsabs.harvard.edu/abs/#3/abstract}%
 {\def\hyper@linkstart##1##2{}%
 \let\hyper@linkend\@empty\citealp[#1][#2]{#3}}}
 \newcommandtwoopt{\citepads}[3][][]{\href{https://ui.adsabs.harvard.edu/abs/#3/abstract}%
 {\def\hyper@linkstart##1##2{}%
 \let\hyper@linkend\@empty\citep[#1][#2]{#3}}}
 \newcommandtwoopt{\citetads}[3][][]{\href{https://ui.adsabs.harvard.edu/abs/#3/abstract}%
 {\def\hyper@linkstart##1##2{}%
 \let\hyper@linkend\@empty\citet[#1][#2]{#3}}}
 \newcommandtwoopt{\citeyearads}[3][][]%
 {\href{https://ui.adsabs.harvard.edu/abs/#3/abstract}
 {\def\hyper@linkstart##1##2{}%
 \let\hyper@linkend\@empty\citeyear[#1][#2]{#3}}}
\makeatother
\begin{document}

   \title{Deep and fast Solar 
       System flybys: \\ The controversial case of WD~0810-353}
   \author{R. de la Fuente Marcos$^{1}$
           \and
           C. de la Fuente Marcos$^{2}$}
   \authorrunning{R. de la Fuente Marcos \and C. de la Fuente Marcos}
   \titlerunning{Is WD~0810-353 a hypervelocity runaway white dwarf?}
   \offprints{R. de la Fuente Marcos, \email{rauldelafuentemarcos@ucm.es}}
   \institute{$^{1}$AEGORA Research Group,
              Facultad de Ciencias Matem\'aticas,
              Universidad Complutense de Madrid,
              Ciudad Universitaria, E-28040 Madrid, Spain \\
              $^{2}$ Universidad Complutense de Madrid,
              Ciudad Universitaria, E-28040 Madrid, Spain}
   \date{Received 20 September 2022 / Accepted 9 October 2022}

% Abstract: 300 words
% Text: 3000 words
   \abstract
      {Most flybys in the Galactic disk are distant, beyond 10$^{4}$~AU, and 
       have characteristic velocities of $\sim$70~km~s$^{-1}$. However, deep and 
       fast encounters also take place, albeit with lower probability, 
       particularly if one of the objects involved is a stellar remnant 
       ejected during a supernova. WD~0810-353 might be a high velocity white 
       dwarf, and it was recently identified as heading straight for the Solar 
       System; however, the \textit{Gaia}~DR3 data that support its future deep 
       and fast flyby are regarded as suspicious.
       }
      {Here, we reanalyze the \textit{Gaia}~DR3 data set associated with
       WD~0810-353 to confirm or reject the reality of its Solar 
       System flyby and also to investigate its possible runaway status. 
       }
      {We studied the evolution of WD~0810-353 forward in time using $N$-body 
       simulations. We computed the distribution of distances of closest 
       approach and their associated times of perihelion passage. We used a 
       statistical analysis of the kinematics of this object to assess its
       possible hypervelocity. We compared its mean BP/RP spectrum to those 
       of other well-studied white dwarfs.
       }
      {We confirm that WD~0810-353 is headed for the Solar 
       System, but the 
       actual parameters of the encounter depend strongly on its radial 
       velocity. The \textit{Gaia}~DR3 value of $-373.74\pm8.18$~km~s$^{-1}$ 
       is strongly disfavored by our analyses. Its mean BP/RP spectrum 
       suggests a value over ten times higher based on the position of its 
       putative H$\alpha$ line. However, spectral matching using other white 
       dwarfs with non-\textit{Gaia} data indicate a radial velocity in the 
       interval ($-$60,~$-$70)~km~s$^{-1}$. 
       }
      {These results confirm the future flyby of WD~0810-353 near the Solar 
       System, although the relative velocity could be high enough or the 
       minimum approach distance large enough to preclude any significant 
       perturbation on the Oort cloud. 
       }


   \maketitle

   \section{Introduction\label{Intro}}
      Our Sun's journey about the Galactic center leads to stellar flybys in which a star may pass close to the Solar System. In general, stars pass 
      by each other at a comfortable distance (well more than 10$^{3}$ to 10$^{4}$~AU; see for example Fig.~1 in \citealt{2022AJ....163...44H}), 
      although this may not be the case in star-forming regions (see for example \citealt{2022arXiv220709752C}) or the Galactic bulge 
      \citep{2020MNRAS.495.2105M}. On the other hand, relative velocities during stellar flybys could be close to or below 1~km~s$^{-1}$ (see for 
      example Table~4 in \citealt{2019MNRAS.489..951D}) but also as high as 300 to 3000~km~s$^{-1}$ for objects ejected during supernova explosions 
      (see for example \citealt{2012ApJ...750L..39T,2015Sci...347.1126G,2015MNRAS.448L...6T,2019MNRAS.489..420R}), and 3000 to 5000~km~s$^{-1}$ for 
      recoil velocities linked to binary black hole mergers (see for example \citealt{2019MNRAS.482.2132D,2022PhRvL.128s1102V}). 

      The average relative speed between two stars selected at random from a Gaussian distribution is given by $v_{\rm r}=\sqrt{2}\ \sigma_v$, where
      $\sigma_v$ is the velocity dispersion (see for example \citealt{2003astro.ph..5512M}). On the other hand, the velocity dispersion of the 
      Galactic thin disk (where the Sun is located) is 48.9~km~s$^{-1}$ \citep{2020AJ....160...43A}, and this translates into a typical relative speed 
      of Solar System flybys of 69.1~km~s$^{-1}$. Therefore, and in theory, the Solar System may experience close encounters with other objects (stars 
      or collapsed objects) at typical velocities under 100~km~s$^{-1}$ (for a representative sample, see, for example, \citealt{2022ApJ...935L...9B}), 
      but flybys at relative velocities close to or below 1~km~s$^{-1}$ (kinematic siblings of the Sun) or above 1000~km~s$^{-1}$ (hypervelocity 
      objects or runaways) are possible as well.

      Using \textit{Gaia}~Data Release 3 (DR3) data, \citet{2022arXiv220614443B} find that the white dwarf WD~0810-353 (also known as UPM~J0812-3529) 
      will approach the Solar System at a distance of 0.150$\pm$0.003~pc in 0.029$\pm$0.001~Myr. This is not one of the most remarkable future Solar 
      System flybys in terms of distance of closest approach, but it certainly is the closest in terms of time of perihelion passage among those 
      already studied. This result has been contested by \citet{2022ApJ...935L...9B}, who argues that the value of the radial velocity of WD~0810-353 
      in \textit{Gaia}~DR3 is probably incorrect because radial velocity determinations for white dwarfs from the \textit{Gaia} pipeline are 
      unreliable. In this paper we reanalyze the \textit{Gaia}~DR3 data set associated with this object to confirm or reject the reality of its future 
      Solar System flyby as discussed by \citet{2022arXiv220614443B} but also to investigate the possible hypervelocity runaway status of this stellar 
      remnant. This paper is organized as follows. In Sect.~\ref{Data} we outline the context of our research, review our methodology, and present 
      the data and tools used in our analyses. The controversial nature of the radial velocity of WD~0810-353 is considered in Sect.~\ref{Controversy}. 
      In Sect.~\ref{Results} we apply our methodology and, in Sect.~\ref{Discussion}, discuss its results. Our conclusions are summarized in 
      Sect.~\ref{Conclusions}.

   \section{Context, methods, and data\label{Data}}
      In the following, we present some theoretical background useful for navigating the reader through the results presented in the sections, the 
      basic details of our approach and the data, and the tools used to obtain the results. 

      \subsection{Context}
         Solar System flybys are  of interest because it is generally thought that sufficiently close stellar flybys can send small bodies hurtling 
         toward the inner Solar System (see for example \citealt{2022A&A...660A.100D}), leading to large-scale impacts on the Earth that may trigger 
         climate change and other harmful large-scale processes (see for example \citealt{2005M&PS...40..817C}). One of the first studies aimed at 
         evaluating the probability and parameters of such events was performed by \citet{1996EM&P...72...19M}. \citet{2018A&A...616A..37B} used 
         \textit{Gaia}~Data Release 2 (DR2) to identify close past and future stellar encounters with the Sun. Using data from \textit{Gaia} and other 
         sources, a number of close stellar encounters with the Solar System have been identified and their characteristic parameters computed (see 
         for example \citealt{2010AstL...36..220B,2010AstL...36..816B,2017AstL...43..559B,2020AstL...46..245B,2021AstL...47..180B,2022arXiv220614443B,
         2020MNRAS.491.2119W,2022A&A...657A..65D,2022A&A...660A.100D,2022A&A...664A.123D}). Past stellar encounters may have perturbed the Oort cloud 
         (see for example \citealt{2018RNAAS...2...30D,2018MNRAS.476L...1D}).

         The existence of hypervelocity stars was first predicted by \citet{1988Natur.331..687H} and later discussed by \citet{2003ApJ...599.1129Y}. 
         WD~0810-353 was first identified as a high proper motion white dwarf by \citet{2018AJ....155..176F}, so this object is not a star in a strict 
         sense but rather a stellar remnant; it was further characterized using \textit{Gaia}~DR2 data by \citet{2019MNRAS.482.4570G}. Hypervelocity 
         runaway white dwarfs and the mechanisms that produce them have only recently been identified (see for example \citealt{2020A&A...641A..52N,
         2021ApJ...923L..34B}).

      \subsection{Methodology}
         The assessment of the future encounter between WD~0810-353 and the Solar System should be based on an analysis of results from a 
         representative sample of $N$-body simulations that take the uncertainties in the input data from \textit{Gaia}~DR3 into account. Here, we 
         carried out such calculations using a direct $N$-body code implemented by \citet{2003gnbs.book.....A} that is publicly available from the 
         website of the Institute of Astronomy of the University of Cambridge.\footnote{\url{https://www.ast.cam.ac.uk/~sverre/web/pages/nbody.htm}} 
         This software uses the Hermite integration scheme as described by \citet{1991ApJ...369..200M}. Results from this code were discussed in 
         detail by \citet{2012MNRAS.427..728D}. The physical model considers perturbations from: the Sun, the four most massive planets, the 
         barycenter of the Pluto-Charon system, and WD~0810-353 (with an assumed mass of 0.63~$M_{\odot}$). This choice is because we are interested 
         in possible effects on Pluto, but Pluto resonant behavior cannot be properly reproduced without Neptune, and Neptune's dynamics cannot be 
         properly simulated without Uranus, Saturn, and Jupiter as the subsystem made of the four giant planets is strongly coupled (see for example 
         \citealt{2007PASJ...59..989T}). Our calculations do not include the Galactic potential as they consist of integrations forward in time for 
         short timescales, while the Sun takes $\sim$220~Myr to complete one revolution around the center of the Galaxy. Our approach has already been 
         used within the context of studying stellar encounters \citep{2018RNAAS...2...30D,2020RNAAS...4..222D,2022RNAAS...6..136D}. 
         
      \subsection{Data, data sources, and tools}
         {\it Gaia}~DR3 \citep{2016A&A...595A...1G,GaiaDR3-2022} provides, among other data, right ascension and declination, absolute stellar 
         parallax, proper motions in right ascension and declination (all referred to epoch 2016.0 or 2457388.5~TDB, Barycentric Dynamical Time), 
         spectroscopic radial velocity, and their respective standard errors for over 3.38$\times$10$^{7}$~sources, all in the solar barycentric 
         reference frame. These data can be transformed into equatorial values as described by \citet{1987AJ.....93..864J}, and state vectors in the 
         ecliptic and mean equinox of reference epoch suitable for Solar System numerical integrations can be computed by applying the usual 
         transformation that involves the obliquity. Here, we used input data from {\it Gaia}~DR3 and barycentric Cartesian state vectors for the 
         Solar System --- provided by Jet Propulsion Laboratory's \textsc{horizons} 
         \citep{2015IAUGA..2256293G},\footnote{\url{https://ssd.jpl.nasa.gov/?horizons}} based on the new DE440/441 planetary ephemeris 
         \citep{2021AJ....161..105P}, and retrieved using resources from the Python package Astroquery \citep{2019AJ....157...98G}. Figures were 
         produced using the Matplotlib library \citep{2007CSE.....9...90H} and statistical tools provided by NumPy \citep{2020Natur.585..357H}.

         In order to compute the  galactocentric Galactic velocity components, we used the software pipeline described by \citet{2019A&A...627A.104D}. 
         Galactocentric positions were found using the value of the distance between the Sun and the Galactic center (Sgr A$^{*}$) given by 
         \citet{2019A&A...625L..10G}, 8.18~kpc. The galactocentric standard of rest is a right-handed coordinate system centered at the Galactic 
         center with positive axes in the directions of the Galactic center (away from it), Galactic rotation, and the North Galactic Pole as 
         discussed by \citet{1987AJ.....93..864J}, for example. Galactocentric Galactic velocity components were calculated as described by 
         \citet{1987AJ.....93..864J}, considering the solar motion values computed by \citet{2010MNRAS.403.1829S} and the value of the in-plane 
         circular motion of the local standard of rest around the Galactic center discussed by \citet{2014ApJ...783..130R}. 

   \section{Caveats of the radial velocity value of WD~0810-353\label{Controversy}}
      In addition to being a putative close approacher in the astronomically near future, WD~0810-353 is a DA magnetic white dwarf with a very strong 
      magnetic field (perhaps as high as 30~MG), a mass of 0.63~$M_{\odot}$, and an age (obtained by interpolation in the cooling curve) of 2.7~Gyr 
      \citep{2020A&A...643A.134B}. In principle, \textit{Gaia}~DR3 \citep{2016A&A...595A...1G,GaiaDR3-2022} provides robust data for this interesting 
      object. WD~0810-353 is designated \textit{Gaia}~DR3~5544743925212648320 and has a parallax of 89.5064$\pm$0.0155~mas, proper motions in right 
      ascension of $-65.479\pm0.016$~mas~yr$^{-1}$ and declination of $-29.204\pm0.018$~mas~yr$^{-1}$, and a radial velocity of 
      $-373.74\pm8.18$~km~s$^{-1}$ with a renormalized unit weight error (RUWE) of 1.039. 

\begin{figure*}
         \begin{center}
            \includegraphics[width=0.49\linewidth]{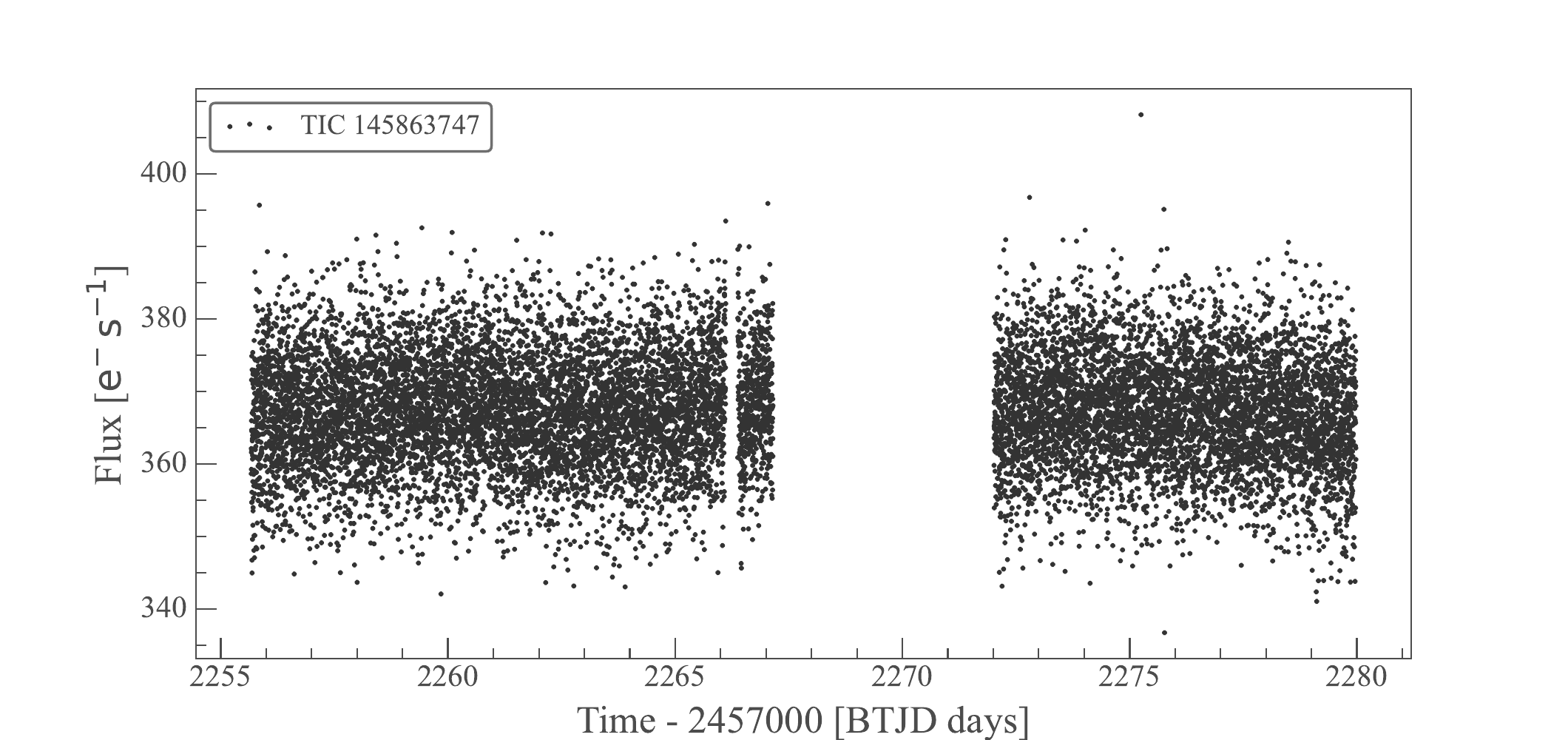}
            \includegraphics[width=0.49\linewidth]{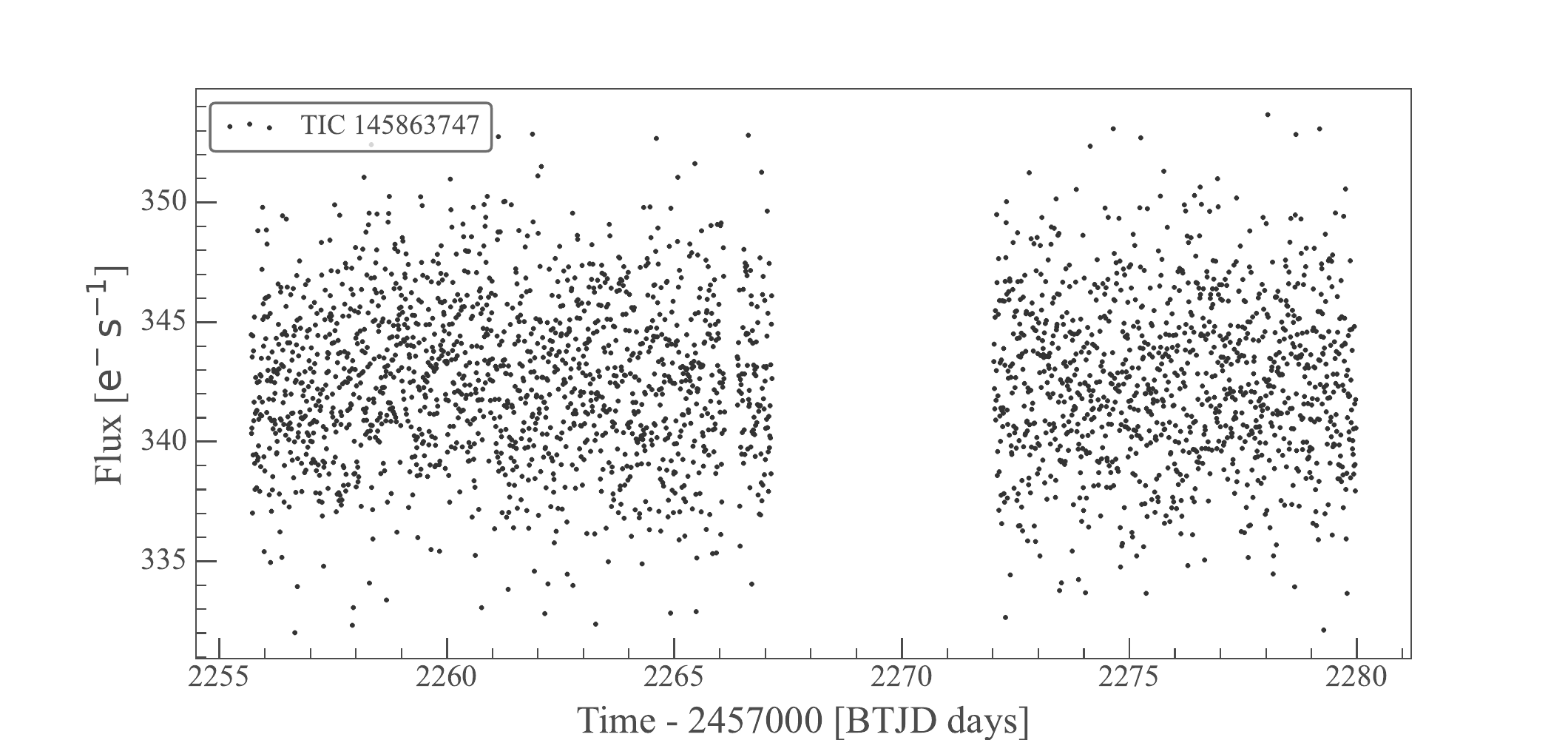}
            \includegraphics[width=0.49\linewidth]{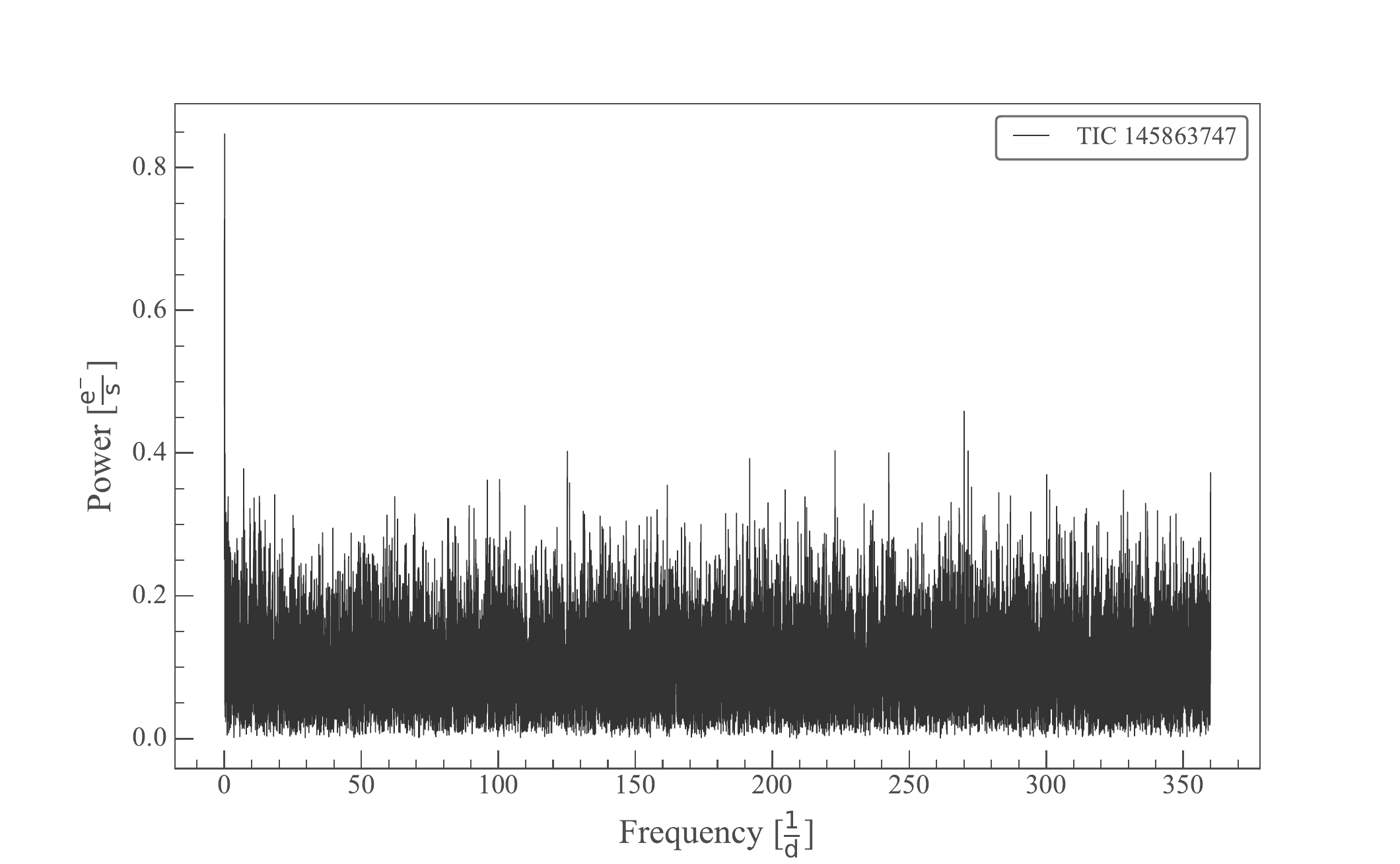}
            \includegraphics[width=0.49\linewidth]{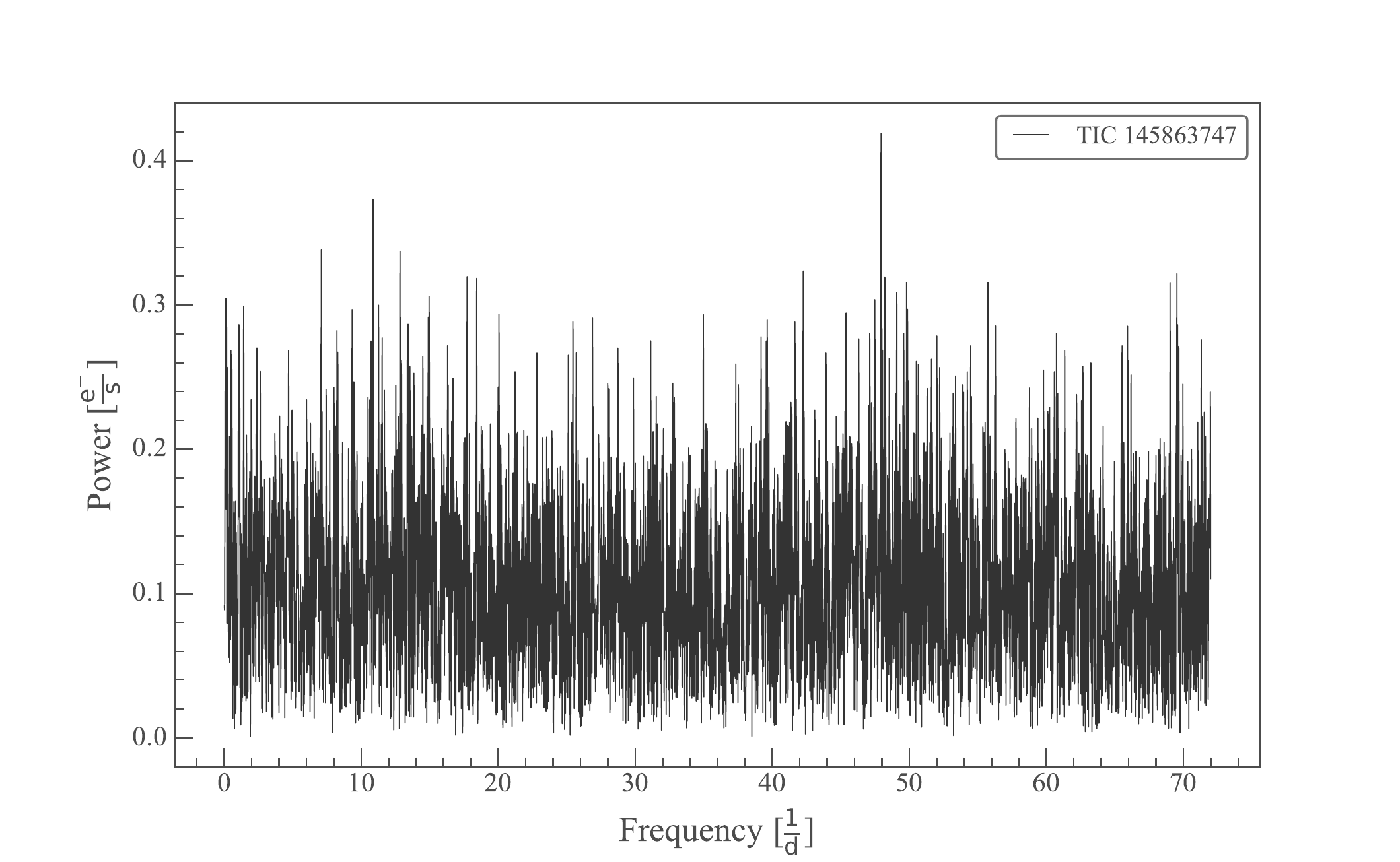}
            \includegraphics[width=0.49\linewidth]{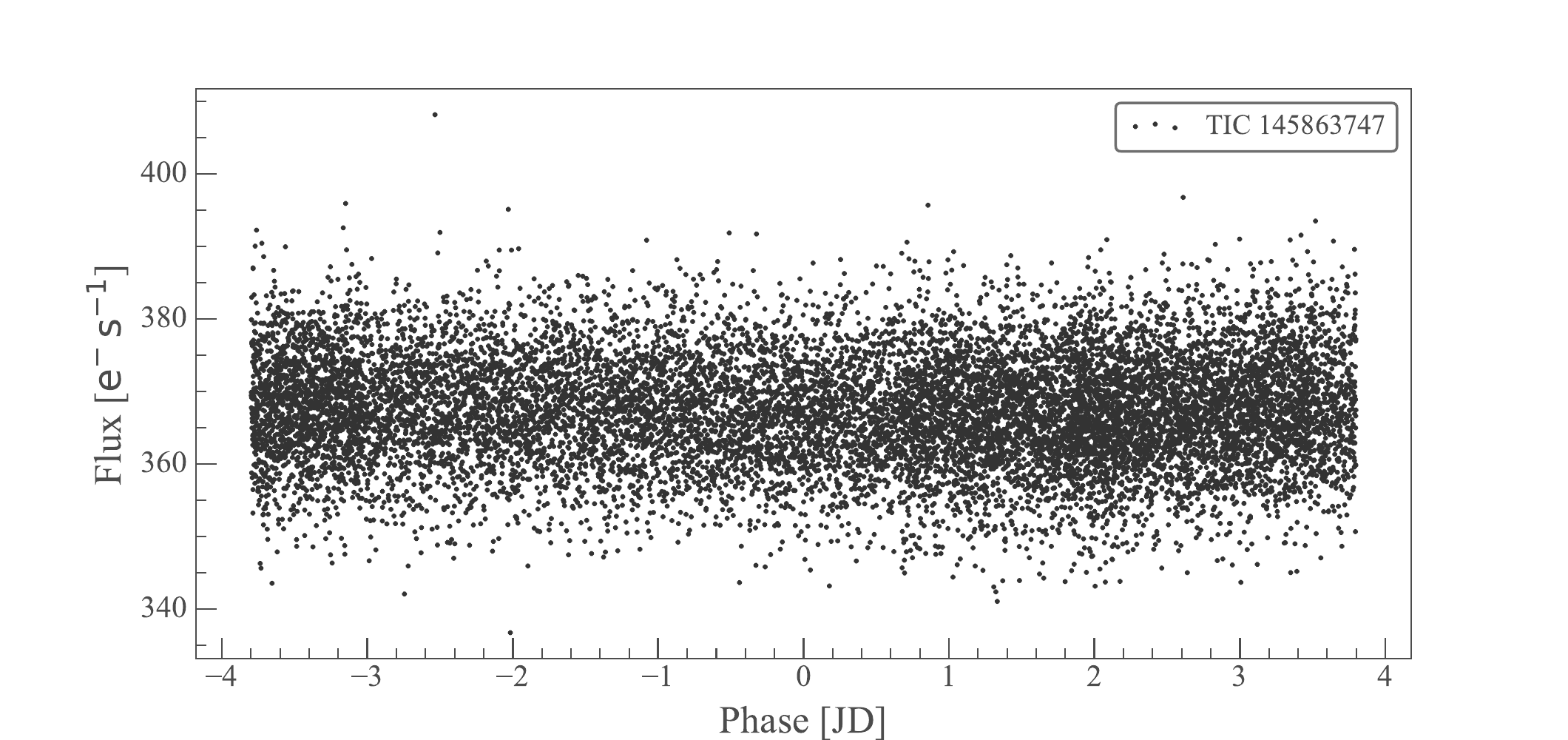}
            \includegraphics[width=0.49\linewidth]{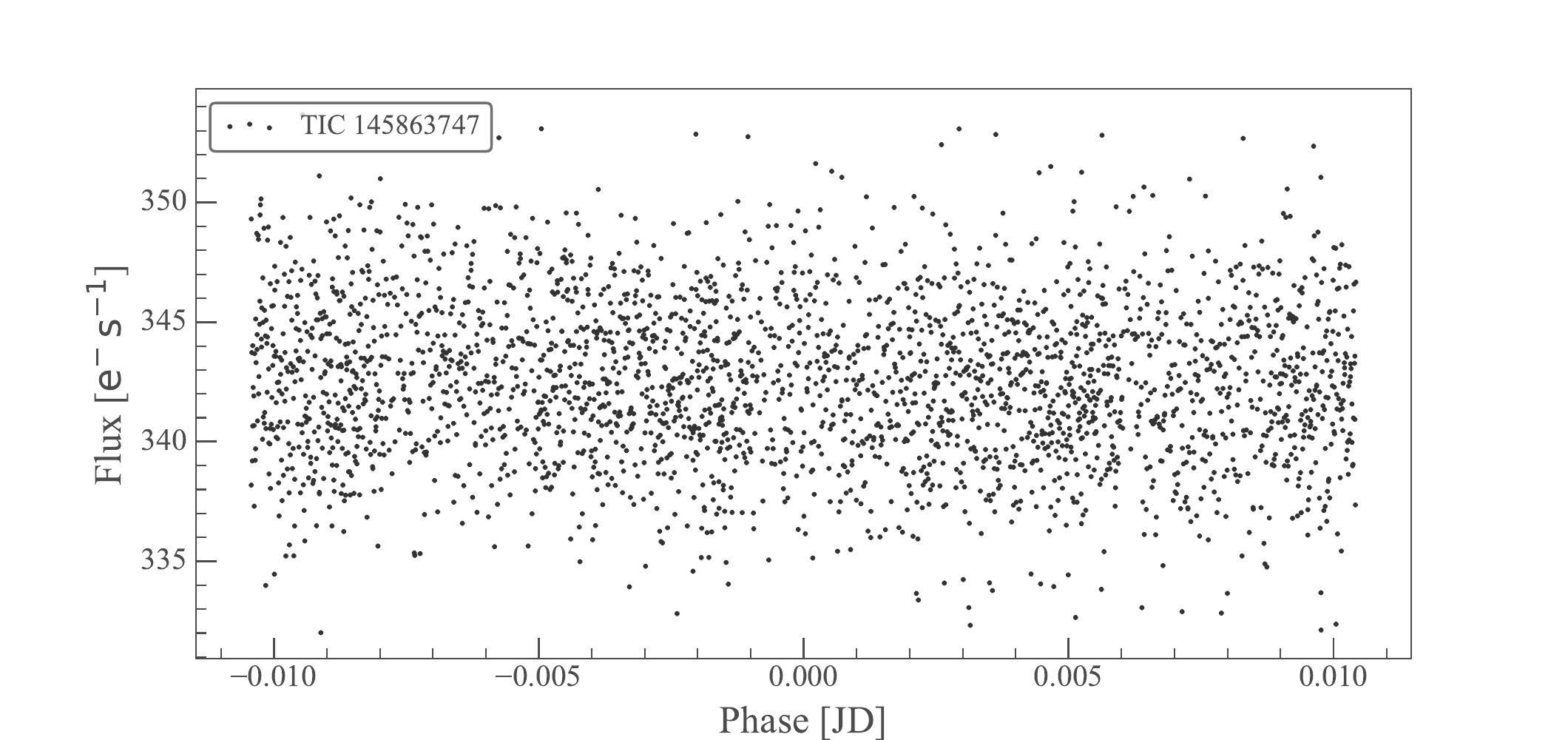}
            \caption{TESS light curve of TIC~145863747 (=WD~0810-353). \textit{Left panels:} Target originally observed at a 120~s cadence. 
                     \textit{Right panels:} In the extended TESS mission the full frame image  cadence is reduced to 600~s. \textit{Top panels:} Light 
                     curve plotted with the tool Lightkurve, a Python package for Kepler and TESS data analysis 
                     \citep{2018ascl.soft12013L,2019AAS...23344501D}. \textit{Middle panels:} Periodogram computed by Lightkurve. 
                     \textit{Bottom panels:} Phased light curve computed by Lightkurve. Both observations include a significant data gap, in each case 
                     caused by the Earth rising above the sunshade on the spacecraft and contributing significant scattered light.
            \label{tess}}
         \end{center}
      \end{figure*}

      If RUWE$>$1.4, the relevant astrometric solution could be problematic and may produce inaccurate 
      results.\footnote{\url{https://gea.esac.esa.int/archive/documentation/GDR2/Gaia_archive/chap_datamodel/sec_dm_main_tables/ssec_dm_ruwe.html}}
      However, \citet{2021ApJ...907L..33S} have pointed out that RUWE values even slightly above 1.0 may signal unresolved binaries in 
      \textit{Gaia} data. WD~0810-353 has also been observed by the Transiting Exoplanet Survey Satellite (TESS) mission \citep{2015JATIS...1a4003R} 
      as TIC~145863747 \citep{2021ApJS..254...39G}. Data from TESS are available at the Mikulski Archive for Space Telescopes 
      (MAST).\footnote{\url{https://archive.stsci.edu/tess/}} The target was observed in Sector 35 in 2021, with light curves extracted from TESS 
      full frame image data. According to light curve data from TESS (see Fig.~\ref{tess}), the object is not variable and as such appears to be a 
      single source, which is consistent with its RUWE of 1.039 from \textit{Gaia}~DR3.
%
%-----------------------------------------------------------------------------------------------------------------------------------------------------
%
%
%-----------------------------------------------------------------------------------------------------------------------------------------------------
%

      The high value of the radial velocity of WD~0810-353 makes it an outlier among the population of known white dwarfs in the solar neighborhood 
      (see for example \citealt{2018MNRAS.480.3942H,2020MNRAS.499.1890M}) and may qualify it  as a member of an even more exclusive class of 
      objects, the hypervelocity runaway white dwarfs. Hypervelocity runaway white dwarfs can be produced as a result of Type Ia supernovae 
      \citep{2017Sci...357..680V}. Survivors are ejected from compact binaries with velocities close to 600~km~s$^{-1}$ or higher, and they might 
      become unbound from their progenitor galaxies \citep{2018ApJ...858....3R,2018MNRAS.479L..96R}. On their way out, they may experience flybys with 
      other stars and their planetary systems, as argued by \citet{2022arXiv220614443B}. In the Milky Way, there are at least four known hypervelocity 
      runaway white dwarfs \citep{2019MNRAS.489.1489R}. However, the critical question here is whether the value of the radial velocity of WD~0810-353 
      in {\it Gaia}~DR3 can be trusted or not.
     
      The Radial Velocity Spectrometer (RVS) is a near-infrared (847--874~nm) instrument with a medium resolution (resolving power of 11\,500); the
      wavelength range works best for G- and K-type stars, and for them radial velocities are derived using three strong ionized calcium lines found 
      at around 849.8, 854.2, and 855.2~nm \citep{2019A&A...622A.205K}. For stars fainter than $G_\mathrm{RVS}$=12~mag (WD~0810-353 has 
      $G_\mathrm{RVS}$=13.67~mag), the {\it Gaia}~DR3 pipeline derives the radial velocity via the cross-correlation of the observed {\it Gaia}-RVS 
      spectra with a synthetic template. Unfortunately, the pipeline does not contain templates for white dwarfs; a template with certain atmospheric 
      parameters (effective temperature, surface gravity, and metallicity) is used instead. For WD~0810-353, $T_\mathrm{eff}$=6000~K, $\log g$=4.5, 
      and $[{\rm Fe/H}]$=$-$1.5~dex were used. It is most likely that this template is inappropriate for a white dwarf.\ The impact of such a template 
      mismatch on the radial velocity has not yet been characterized, but the use of inconsistent templates for other object types, for example 
      emission line stars, has led to systematic errors of several hundred~km~s$^{-1}$ \citep{2022arXiv220605902K}.
  
\begin{figure}
        \centering
         \includegraphics[width=\linewidth]{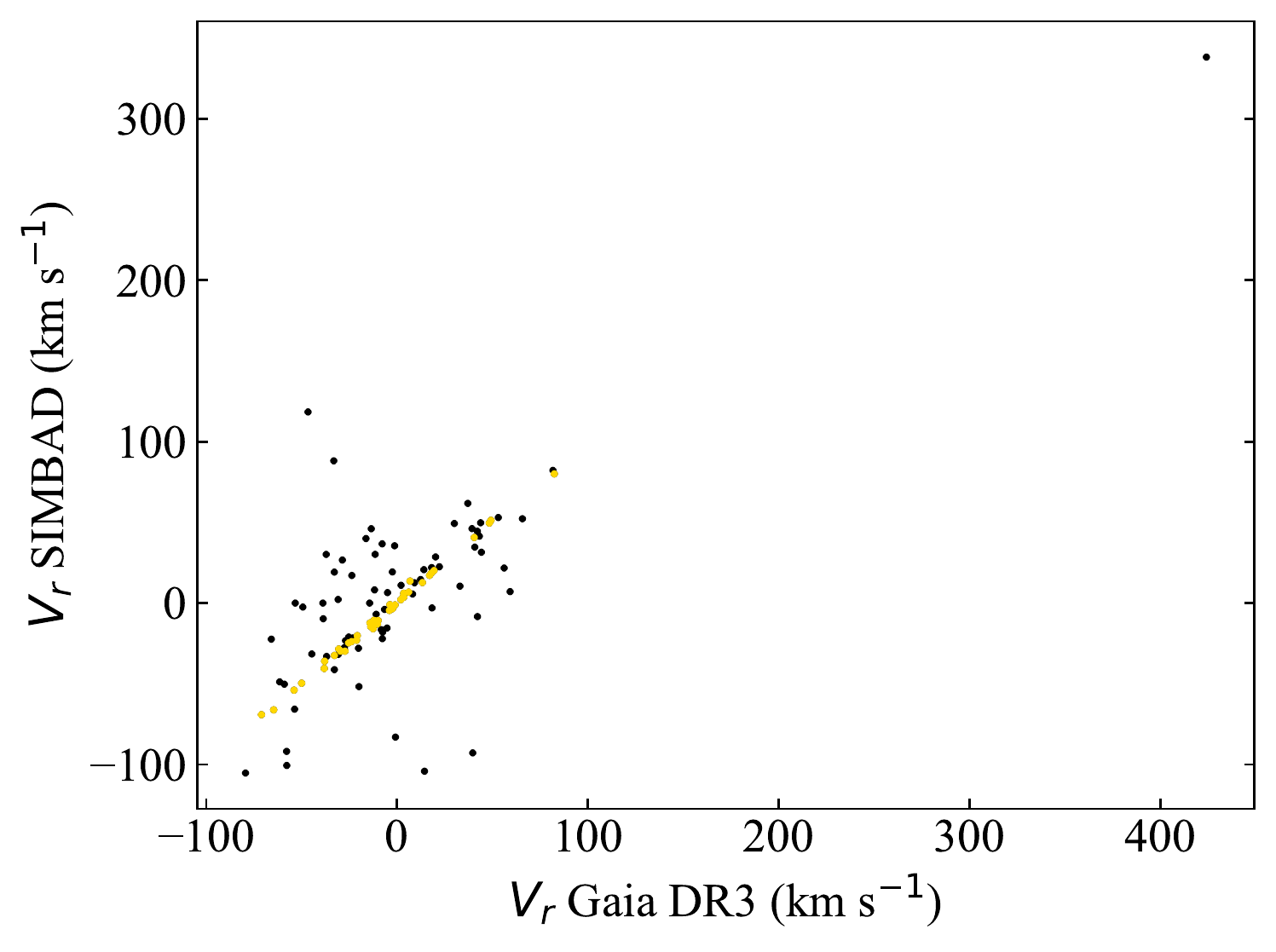}
         \includegraphics[width=\linewidth]{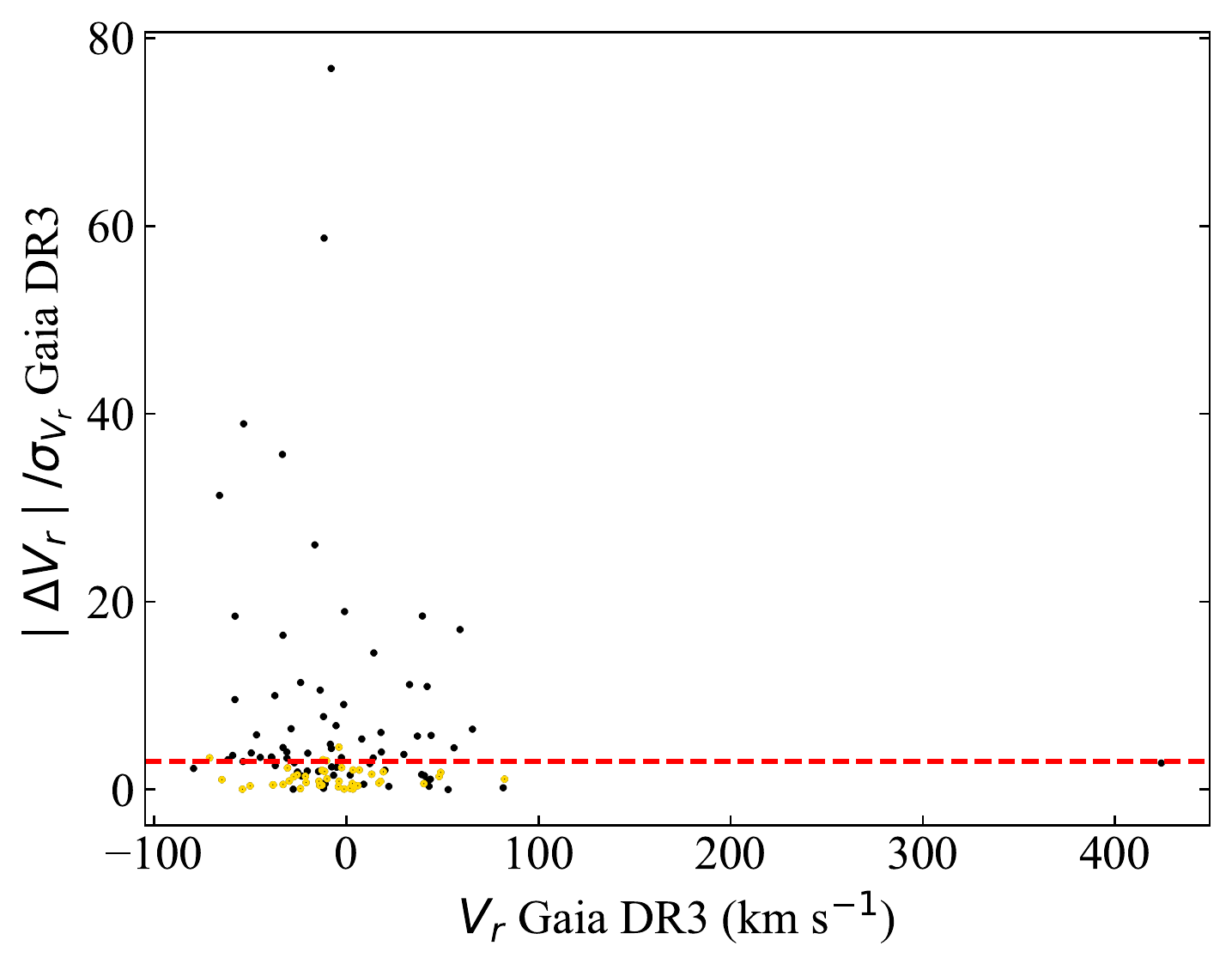}
         \caption{Correlation between radial velocity values of white dwarfs in SIMBAD and {\it Gaia}~DR3. \textit{Top panel:} SIMBAD radial velocity 
                  as a function of the corresponding {\it Gaia}~DR3 value. \textit{Bottom panel:} Absolute value of the difference between the SIMBAD 
                  and the {\it Gaia}~DR3 values divided by the uncertainty in the {\it Gaia}~DR3 value. The dashed red line signals the 3$\sigma$ 
                  difference. Points in gold signal SIMBAD values from {\it Gaia}~DR2.  
                 }
         \label{SIMBAD}
      \end{figure}

      In order to quantify the impact of using inadequate templates to compute the radial velocities of white dwarfs in {\it Gaia}~DR3, we 
      investigated how many white dwarfs have {\it Gaia}~DR3 radial velocities that are in agreement with non-{\it Gaia} values from the literature. 
      This analysis would not unambiguously prove that the specific radial velocity of WD~0810-353 is valid, but would provide reassurances regarding 
      the ability of the {\it Gaia}~DR3 pipeline to measure the radial velocities of some white dwarfs despite the lack of an appropriate template in 
      the pipeline. Our master list of white dwarfs was collected from the SIMBAD\footnote{\url{https://simbad.cds.unistra.fr/simbad/}} astronomical 
      database \citep{2000A&AS..143....9W} using the Simbad module from the Python package Astroquery to query the Simbad service. There are 44\,457 
      sources in SIMBAD catalogued as white dwarfs. Out of this sample, 19\,638 have radial velocities from the literature (some but not many of them 
      from {\it Gaia} DR2). Out of this smaller sample, 114 have both radial velocity values in SIMBAD (41 from {\it Gaia}~DR2) and {\it Gaia}~DR3. 
      The correlation between the two sets of values for this smaller sample is shown in Fig.~\ref{SIMBAD}. Therefore, it is true that some 
      non-{\it Gaia} radial velocities correlate well with their associated {\it Gaia}~DR3 values, but, in general, the correlation is not 
      statistically significant. However, for the outlier white dwarf LB~3209, the SIMBAD value is +338.3$\pm$2.1~km~s$^{-1}$ and the {\it Gaia}~DR3 
      measurement is +424$\pm$30~km~s$^{-1}$, which means that there is a marginal agreement between the two values at the 2.9$\sigma$ level. 
%
%-----------------------------------------------------------------------------------------------------------------------------------------------------
%
%
%-----------------------------------------------------------------------------------------------------------------------------------------------------
%
      
\begin{figure}
        \centering
         \includegraphics[width=\linewidth]{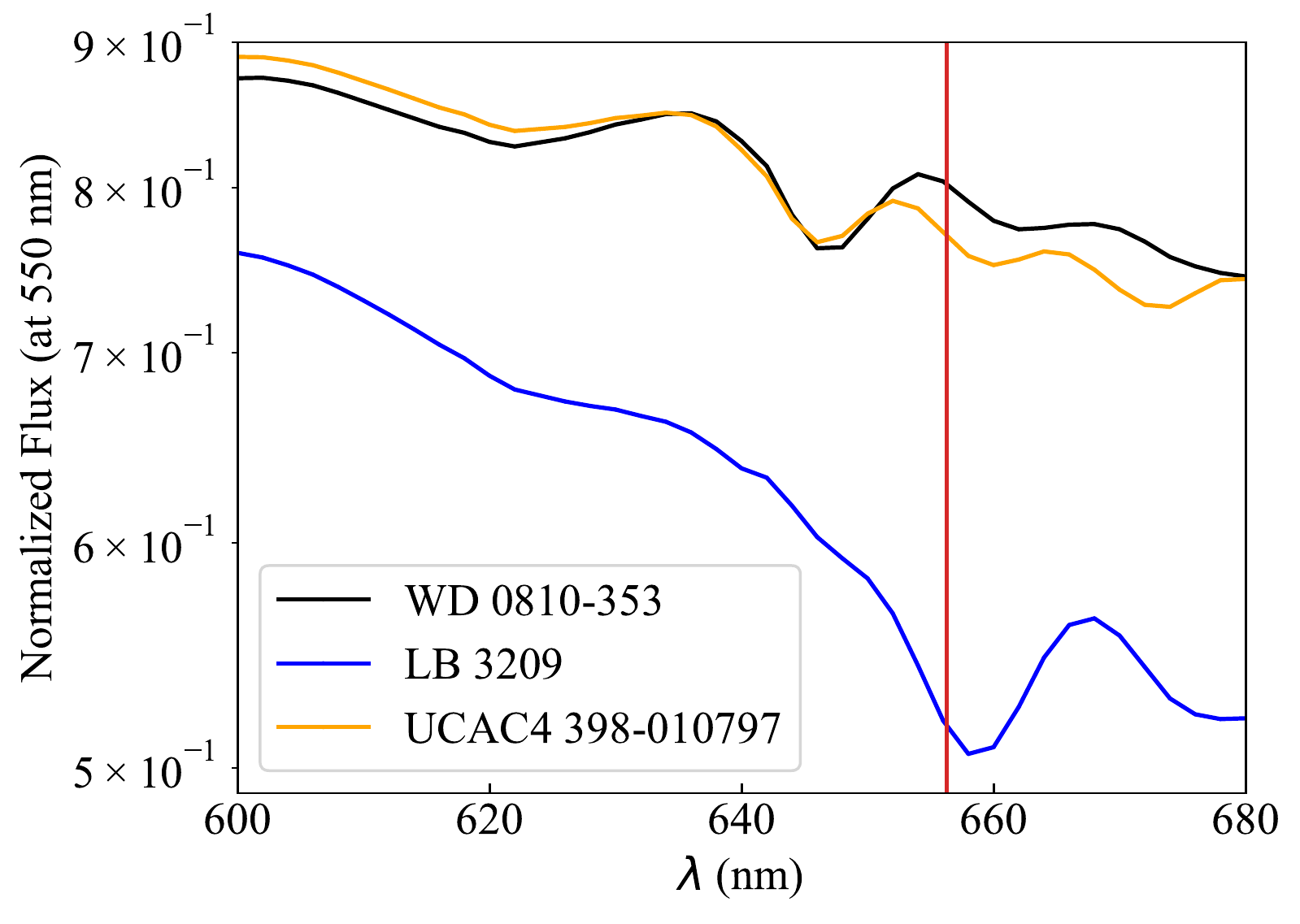}
         \caption{Low-resolution \textit{Gaia}~DR3 mean BP/RP spectra of WD~0810-353, LB~3209, and UCAC4~398-010797. For reference, the vertical red
                  line marks the position of the H$\alpha$ line at rest (656.3~nm). 
                 }
         \label{spectraXP}
      \end{figure}

      The {\it Gaia}~DR3 pipeline computes radial velocities by studying the positions of one or more spectral lines. In the case of white dwarfs, 
      their rather flat spectra provide few lines useful for this task. One of them is the 656.3 nm (first Balmer, in air) line of hydrogen 
      (H$\alpha$). In addition to astrometry and radial velocities, \textit{Gaia}~DR3 provides BP/RP externally calibrated sampled mean spectra (the 
      low-resolution XP spectra in the wavelength range 330--1050~nm) for 219\,197\,643 sources \citep{2022arXiv220606143D,2022arXiv220606205M}. 
      WD~0810-353 and LB~3209 are part of this sample. Figure~\ref{spectraXP} shows the \textit{Gaia}~DR3 spectra (normalized so they all have 1 as 
      the relative flux at 550~nm) of three white dwarfs: WD~0810-353, LB~3209, and UCAC4~398-010797. For LB~3209 (RUWE=0.951), the H$\alpha$ appears 
      redshifted and its position is consistent with that expected for the SIMBAD value of +338.3$\pm$2.1~km~s$^{-1}$. White dwarf LB~3209 is the 
      conspicuous outlier in Fig.~\ref{SIMBAD}. On the other hand, if the absorption feature observed in the spectrum of WD~0810-353 at about 647~nm 
      is a blueshifted H$\alpha$, then the associated radial velocity is close to $-$4300~km~s$^{-1}$. If this interpretation is correct, the 
      \textit{Gaia}~DR3 value of $-373.74\pm8.18$~km~s$^{-1}$ is wrong. We have to emphasize here that the resolving power of the \textit{Gaia}~DR3 XP 
      spectra (not the RVS spectra) is regarded as too low to measure a radial velocity that could be compared to values from the literature. However, 
      the case of LB~3209 is intriguing as the RVS value, the SIMBAD value, and the one derived from the XP spectrum are all somewhat consistent.
%
%-----------------------------------------------------------------------------------------------------------------------------------------------------
%
%
%-----------------------------------------------------------------------------------------------------------------------------------------------------
%

      Figure~\ref{spectraXP} also shows the low-resolution \textit{Gaia}~DR3 mean BP/RP spectrum of UCAC4~398-010797 (RUWE=1.053), which seems to be a 
      very close match for WD~0810-353 in the displayed wavelength interval. White dwarf UCAC4~398-010797 also has a high proper motion in 
      \textit{Gaia}~DR3 (proper motions in right ascension of $-64.18\pm0.02$~mas~yr$^{-1}$ and declination of $-152.84\pm0.02$~mas~yr$^{-1}$), but it 
      has no radial velocity. The spectrum of UCAC4~398-010797 tells us that WD~0810-353 is not unique and that if the absorption feature in the
      spectrum of WD~0810-353 is a blueshifted H$\alpha$ then UCAC4~398-010797 may also be approaching at a speed close to 4000~km~s$^{-1}$. It is
      certainly very unusual to find one object with such a high speed approaching the Solar System, but finding two such objects strongly suggests 
      that the absorption feature discussed above may not be a blueshifted H$\alpha$.

\begin{figure}
        \centering
         \includegraphics[width=\linewidth]{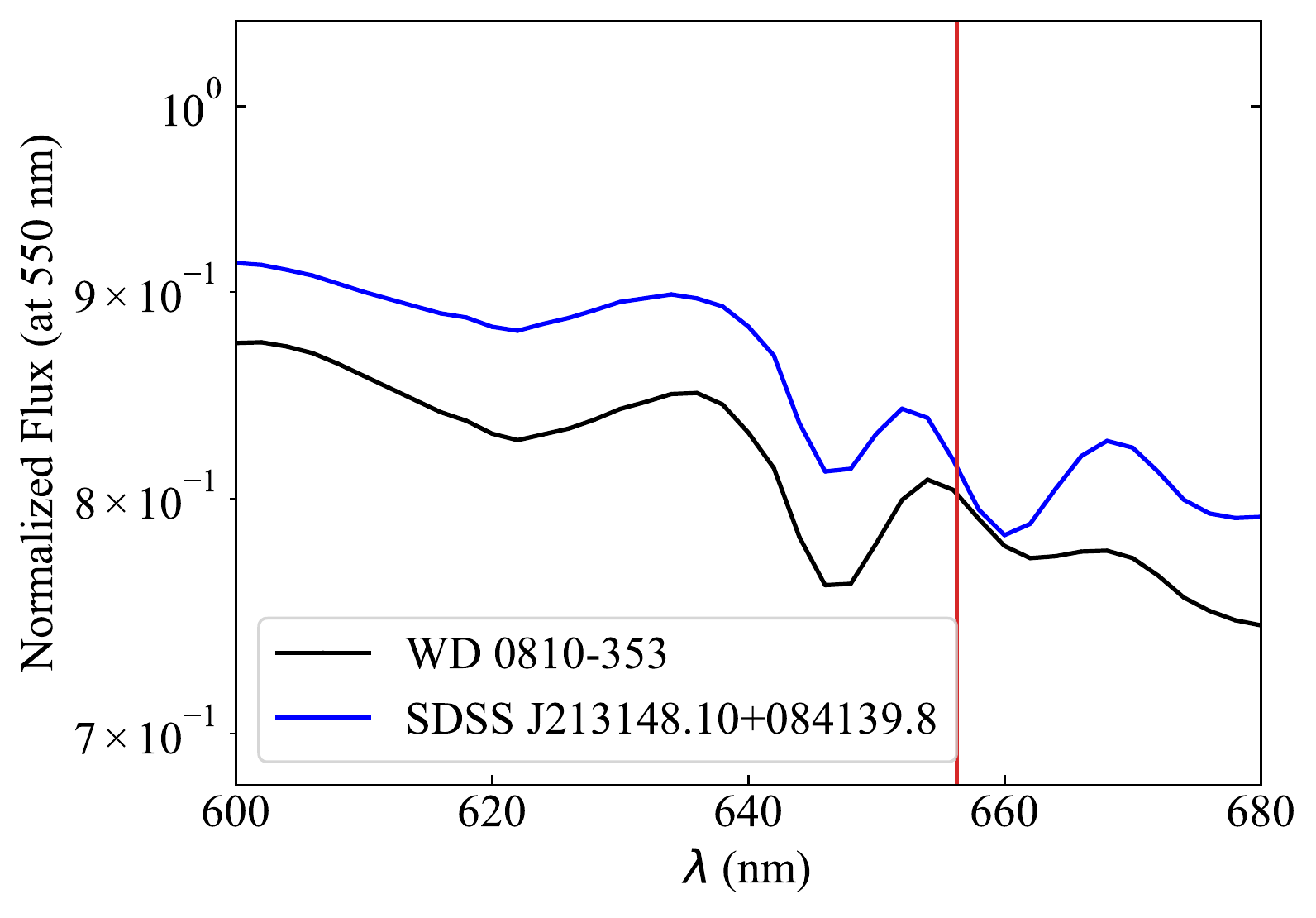}
         \includegraphics[width=\linewidth]{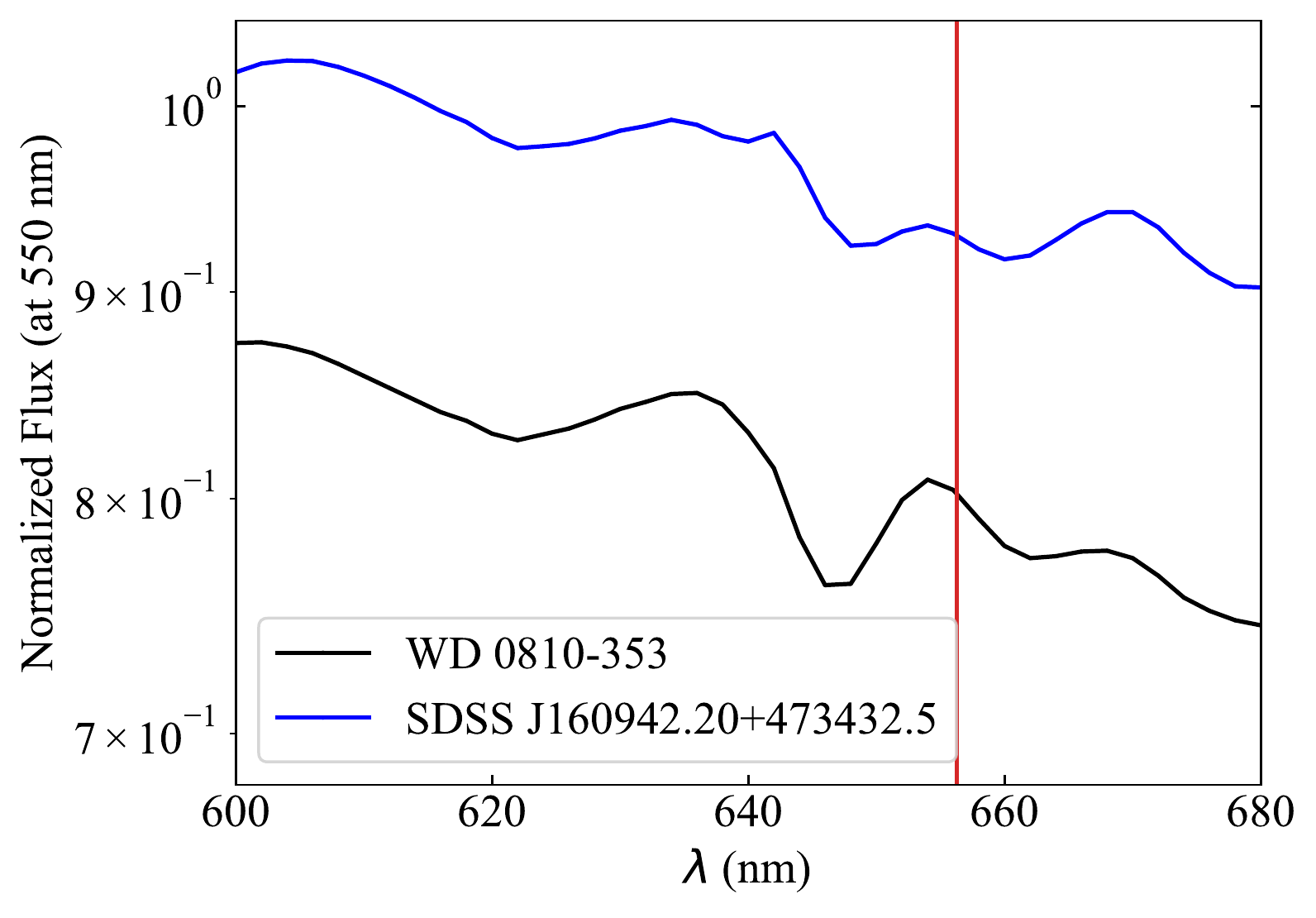}
         \caption{Low-resolution \textit{Gaia}~DR3 mean BP/RP spectra of spectral analogs of WD~0810-353. \textit{Top panel:} SDSS~J213148.10+084139.8.
                  \textit{Bottom panel:} SDSS~J160942.20+473432.5. For reference, the vertical red line marks the position of the H$\alpha$ line at 
                  rest (656.3~nm). 
                 }
         \label{spectraXPref}
      \end{figure}

      In order to further clarify the status of objects such as WD~0810-353 and UCAC4~398-010797, we studied the sample of white dwarfs with a 
      non-{\it Gaia} radial velocity in SIMBAD and the XP spectrum that resembles that of WD~0810-353. Two such objects are SDSS~J213148.10+084139.8 
      and SDSS~J160942.20+473432.5 (see Fig.~\ref{spectraXPref}). White dwarf SDSS~J213148.10+084139.8 has a radial velocity in SIMBAD of 
      $-$63$\pm$3~km~s$^{-1}$ and $-$39$\pm$11~km~s$^{-1}$ in \textit{Gaia}~DR3; SDSS~J160942.20+473432.5 has respective values of 
      $-$23$\pm$2~km~s$^{-1}$ and $-$21.3$\pm$1.2~km~s$^{-1}$. For SDSS~J213148.10+084139.8 the two values are consistent at the 2.2$\sigma$ level; 
      for SDSS~J160942.20+473432.5 consistency is at the 1.4$\sigma$ level. Both cases suggest that the absorption feature in the spectrum of 
      WD~0810-353 is not H$\alpha$ and that the actual radial velocity of WD~0810-353 could be a relatively normal value well below $-$100~km~s$^{-1}$.    
%
%-----------------------------------------------------------------------------------------------------------------------------------------------------
%
%
%-----------------------------------------------------------------------------------------------------------------------------------------------------
%

   \section{Results\label{Results}} 
      Here, we first used \textit{Gaia}~DR3 data to investigate the possible hypervelocity runaway nature of this stellar remnant and to reproduce the
      details of the future flyby of WD~0810-353 near the Solar System as discussed by \citet{2022arXiv220614443B}. However, and as pointed out above, 
      the value of the radial velocity of WD~0810-353 in \textit{Gaia}~DR3 could be incorrect, and alternative scenarios, namely the extreme 
      hypervelocity case and one standard flyby, are discussed as well.  

      \subsection{Hypervelocity runaway white dwarf}
         Using the approach described above, we computed both the galactocentric distance and velocity for WD~0810-353; we also found the angle between 
         the position and velocity vectors. As is customary in \textit{Gaia}-related publications, we provide the median (50th percentile) and the 
         16th and 84th percentiles (as the quality of the published data is very good, the dispersion is small and symmetrical). The galactocentric 
         velocity is $v_{\rm GC}=617\pm8$~km~s$^{-1}$, and the angle is $\theta=101.16\degr\pm0.10$\degr; the galactocentric Galactic velocity 
         components are $U=119\pm2$~km~s$^{-1}$, $V=605\pm8$~km~s$^{-1}$, and $W=8.7\pm0.4$~km~s$^{-1}$. \citet{2019MNRAS.485.3514D} find a value for 
         the escape speed in the neighborhood of the Sun of $528^{+24}_{-25}$~km~s$^{-1}$. Therefore, and for the value of the radial velocity in 
         \textit{Gaia}~DR3, WD~0810-353 may already be unbound to the Milky Way at the 3.6$\sigma$ level. If we consider the effect of Einstein's 
         gravitational redshift and perform the correction pointed out above, the values are: $v_{\rm GC}=667\pm8$~km~s$^{-1}$, 
         $\theta=101.59\degr\pm0.09$\degr, $U=133\pm2$~km~s$^{-1}$, $V=653\pm8$~km~s$^{-1}$, and $W=9.4\pm0.4$~km~s$^{-1}$. WD~0810-353 is moving 
         tangentially, and the value of the angle is in the neighborhood of 90{\degr} as both vectors are nearly perpendicular, which is often 
         interpreted as suggestive of an extragalactic origin if unbound. However, the relatively low estimated age of WD~0810-353 as given by 
         \citet{2020A&A...643A.134B} helps in discarding a possible intergalactic provenance. The trajectory has a small inclination with respect to 
         the disk, so it was probably ejected from the thick or even the thin disk. 

      \subsection{Solar System flyby}
\begin{figure*}
         \begin{center}
            \includegraphics[scale=0.6,angle=0]{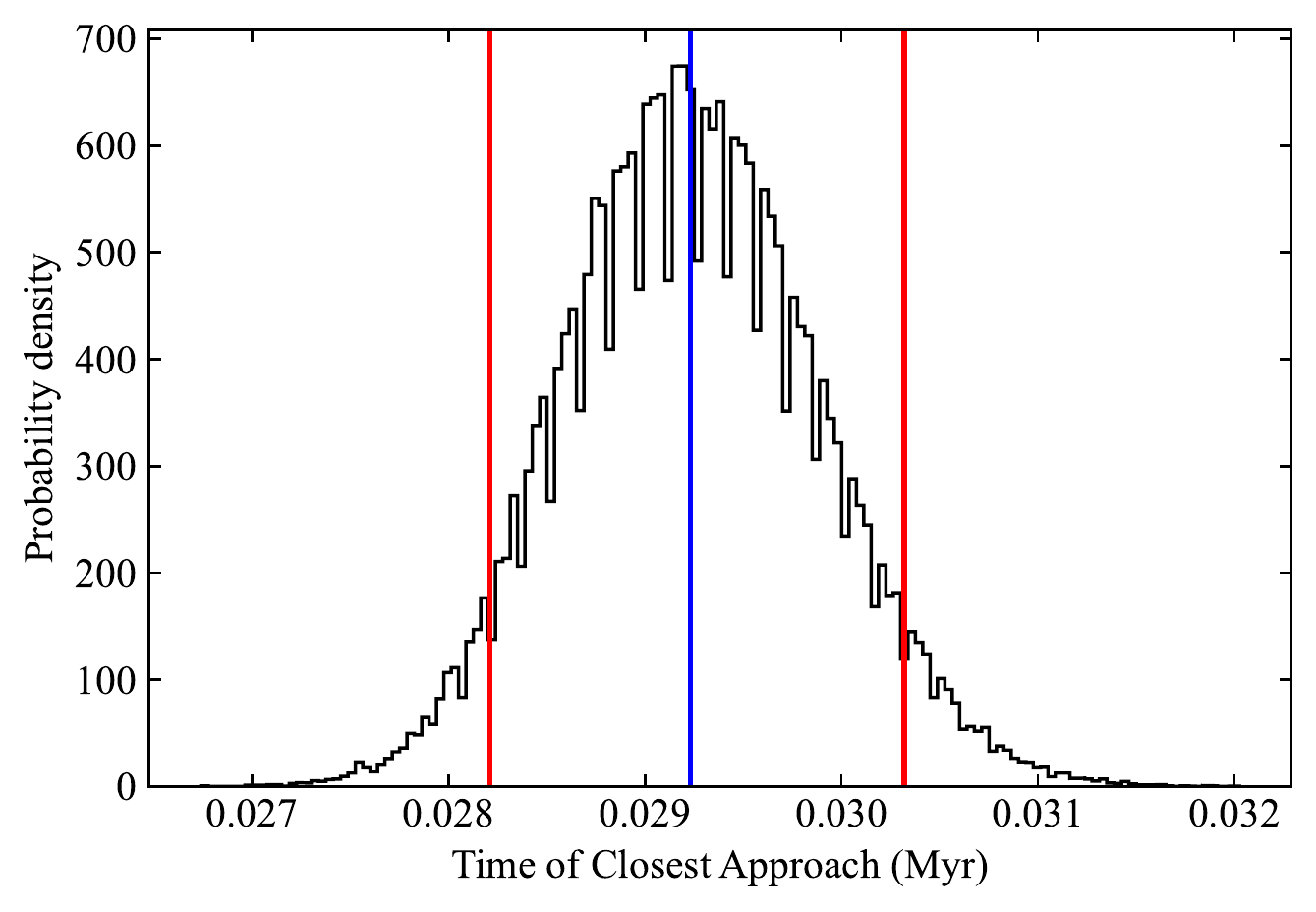}
            \includegraphics[scale=0.6,angle=0]{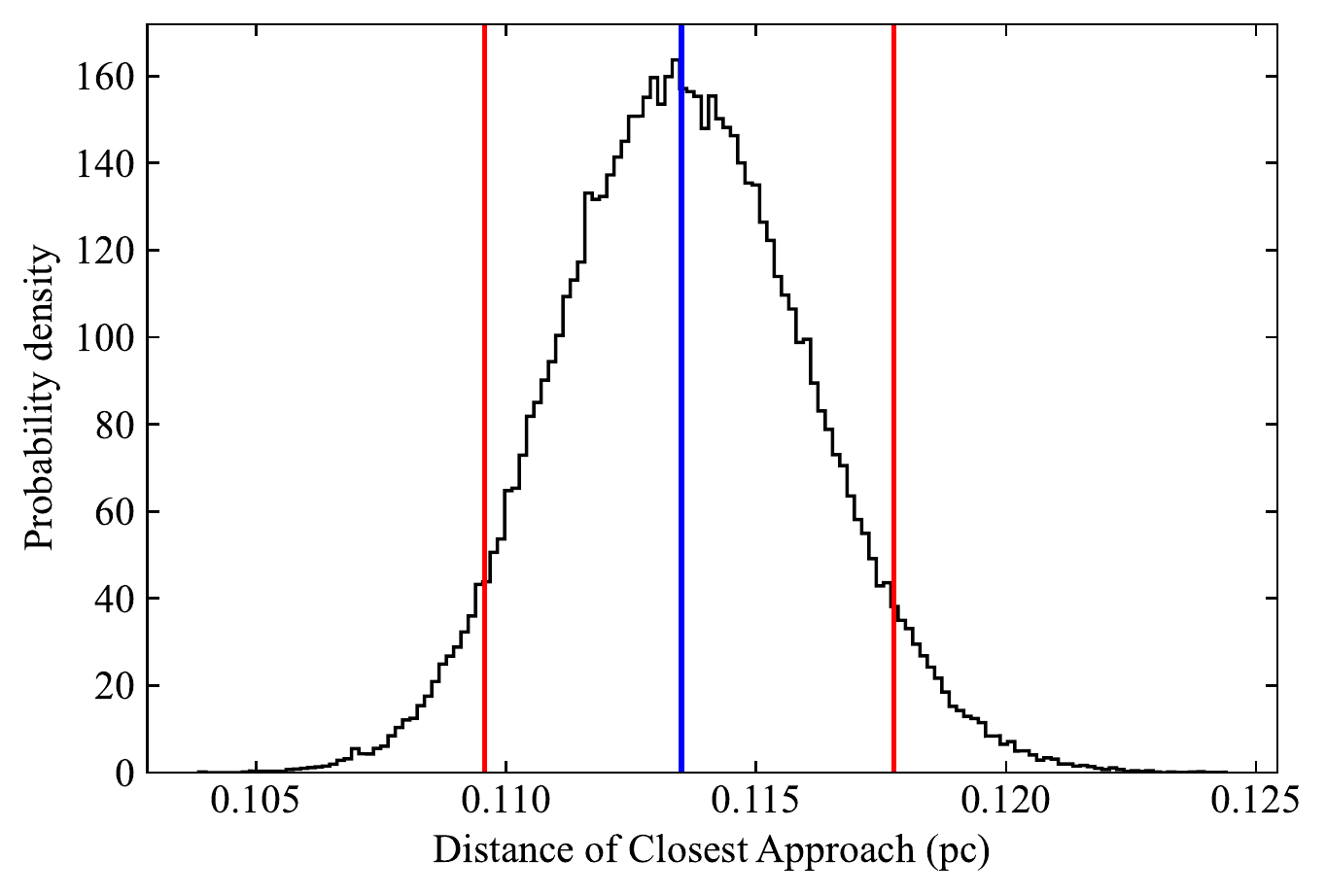}
            \includegraphics[scale=0.6,angle=0]{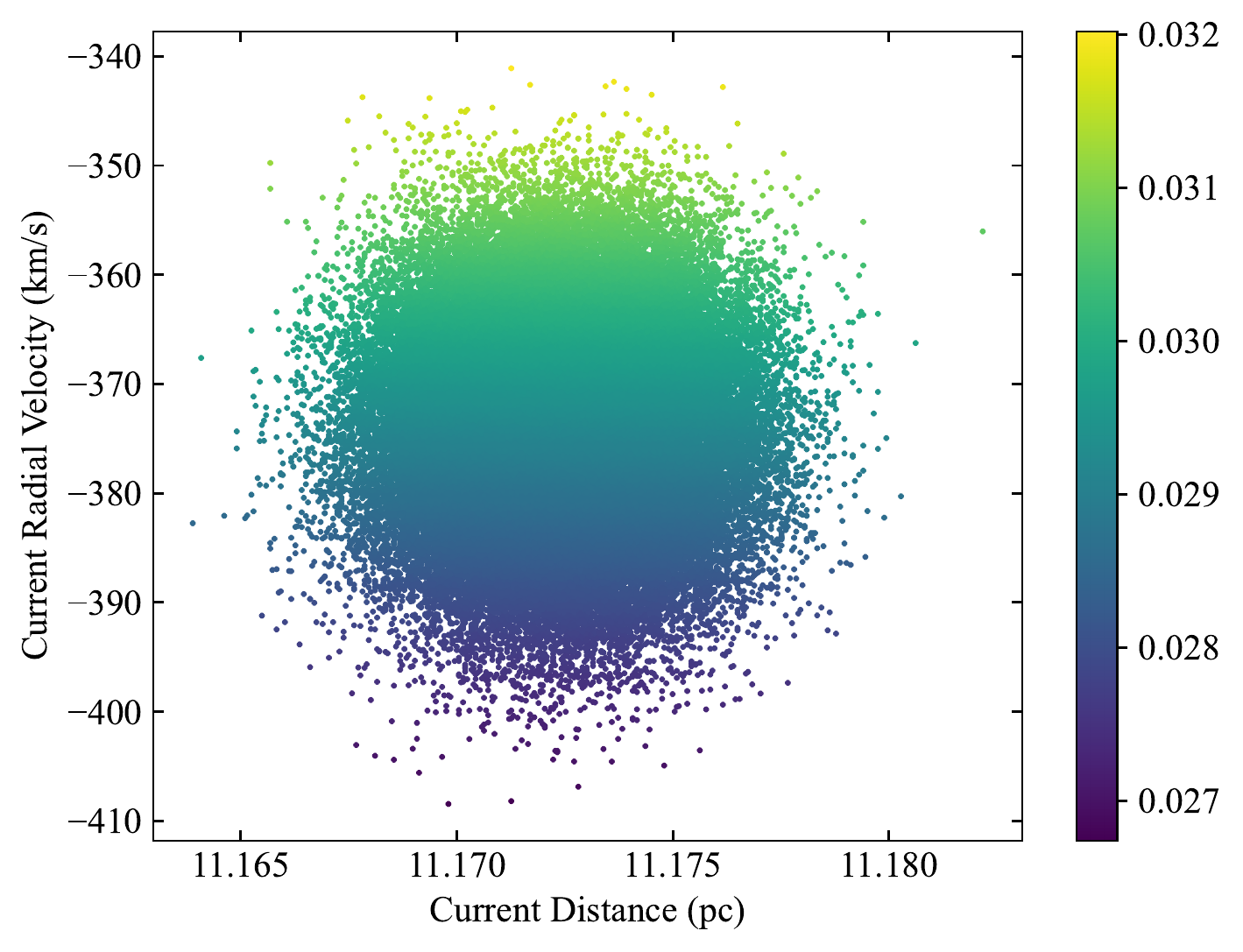}
            \includegraphics[scale=0.6,angle=0]{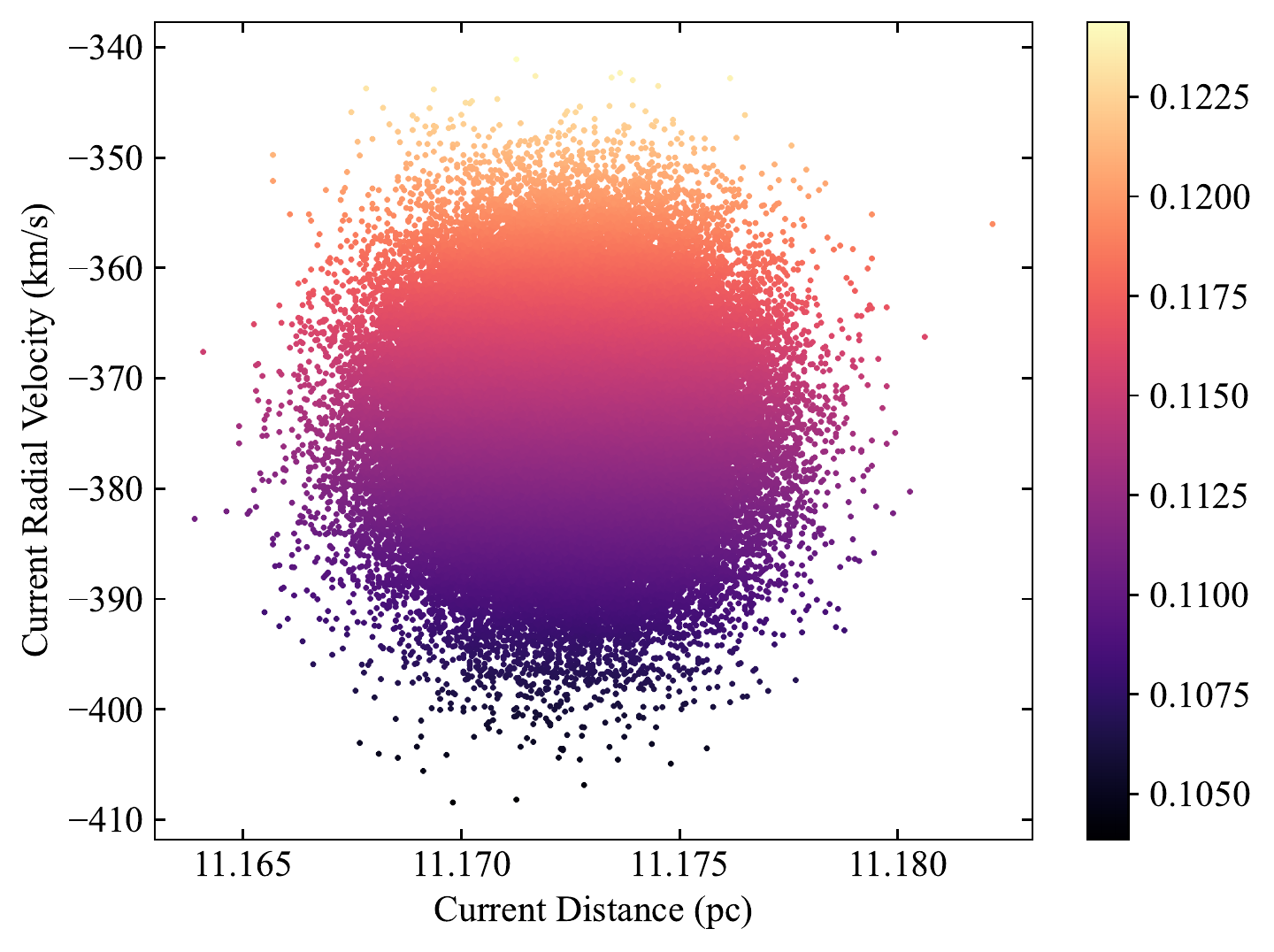}
            \caption{Future perihelion passage of WD~0810-353 as estimated from \textit{Gaia}~DR3 input data and the $N$-body simulations discussed in 
                     the text without accounting for the effect of Einstein's gravitational redshift. The distribution of times of perihelion passage 
                     is shown in the top-left panel and perihelion distances in the top-right one. The blue vertical lines mark the median values, and 
                     the red ones show the 5th and 95th percentiles. The bottom panels show the times of perihelion passage in megayears (bottom left) 
                     and the distance of closest approach in parsecs (bottom right) as a function of the observed values of the radial velocity of 
                     WD~0810-353 and its distance (randomly generated using the mean values and standard deviations from \textit{Gaia}~DR3), both as 
                     color-coded scatter plots of the distribution in the associated top panel. Histograms were produced using the Matplotlib library 
                     \citep{2007CSE.....9...90H} with sets of bins computed using Astropy \citep{2013A&A...558A..33A,2018AJ....156..123A} by applying 
                     the Freedman and Diaconis rule \citep{FD81}; instead of considering frequency-based histograms, we used counts to form a 
                     probability density such that the area under the histogram sums to one. The color map scatter plot was also produced using 
                     Matplotlib.
            \label{flyby}}
         \end{center}
      \end{figure*}

         Figure~\ref{flyby} shows the results of 10$^{5}$ integrations of control orbits of WD~0810-353 generated using input data from 
         \textit{Gaia}~DR3. In order to confirm or reject the results obtained by \citet{2022arXiv220614443B}, we focused on providing an estimate of 
         the value of the distance of closest approach and its associated time of perihelion passage. The distribution of times is shown in the 
         top-left panel: the mean and standard deviation are 0.0292$\pm$0.0006~Myr, the median value is 0.0292~Myr, and 0.0282--0.0303~Myr is the 90\% 
         confidence interval. The distribution of distances of closest approach in the top-right panel has a mean and standard deviation of 
         0.114$\pm$0.002~pc (or 23\,424$\pm$514~AU, median of 23\,412~AU) with a 90\% probability of coming within 0.110--0.118~pc ($<$0.120~pc, 
         99\%).\ This places WD~0810-353 well inside the Oort cloud (see for example \citealt{1950BAN....11...91O}). The bottom panels of 
         Fig.~\ref{flyby} show that faster approaches tend to produce earlier and closer flybys. However, the effects of the flyby on the orbit of the 
         Pluto-Charon system (and therefore, on the classical trans-Neptunian belt) are still negligible. Our results are consistent with those in 
         \citet{2022arXiv220614443B} and \citet{2022ApJ...935L...9B}.
%
%-----------------------------------------------------------------------------------------------------------------------------------------------------
%
%
%-----------------------------------------------------------------------------------------------------------------------------------------------------
%

\begin{figure*}
         \begin{center}
            \includegraphics[scale=0.6,angle=0]{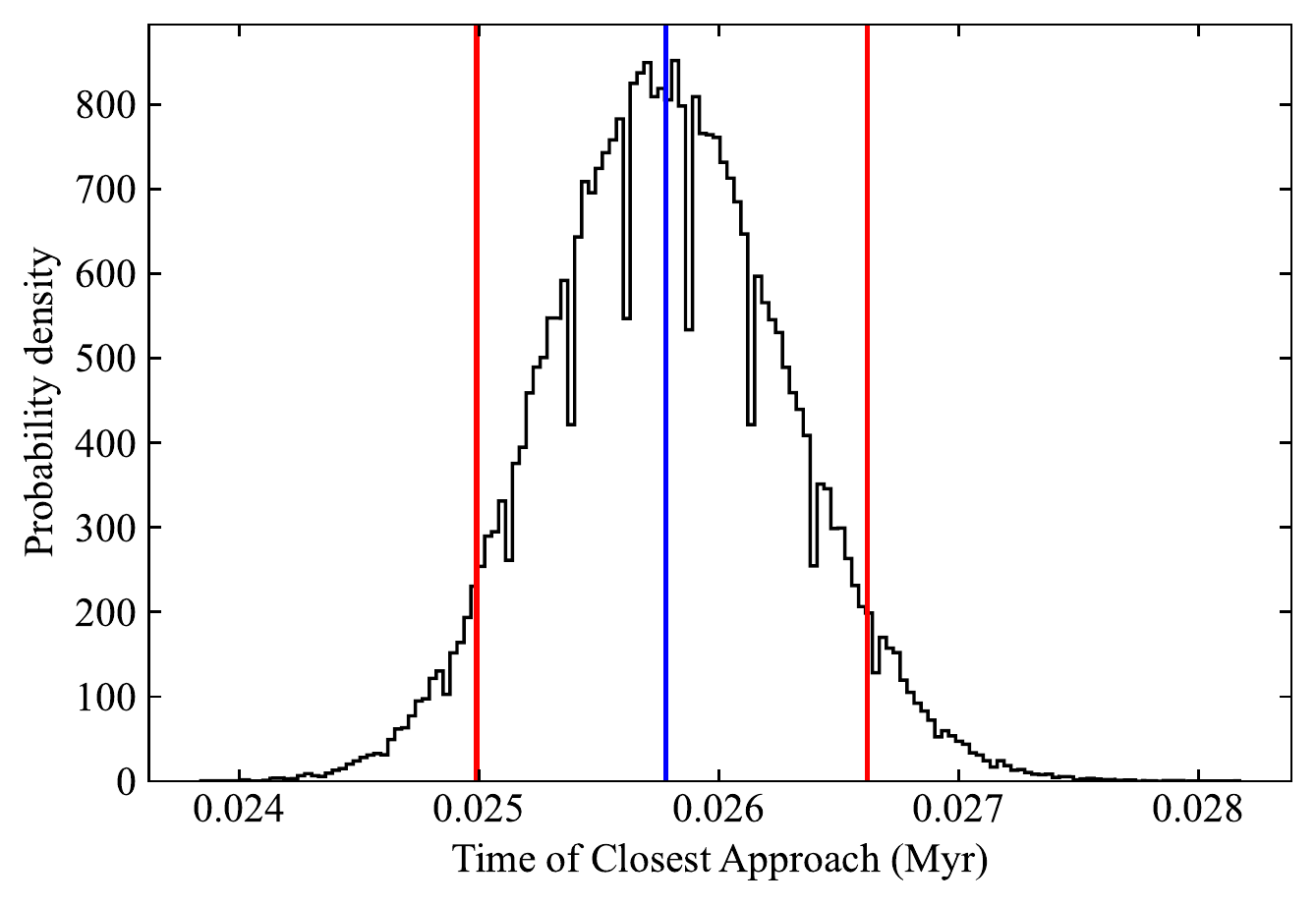}
            \includegraphics[scale=0.6,angle=0]{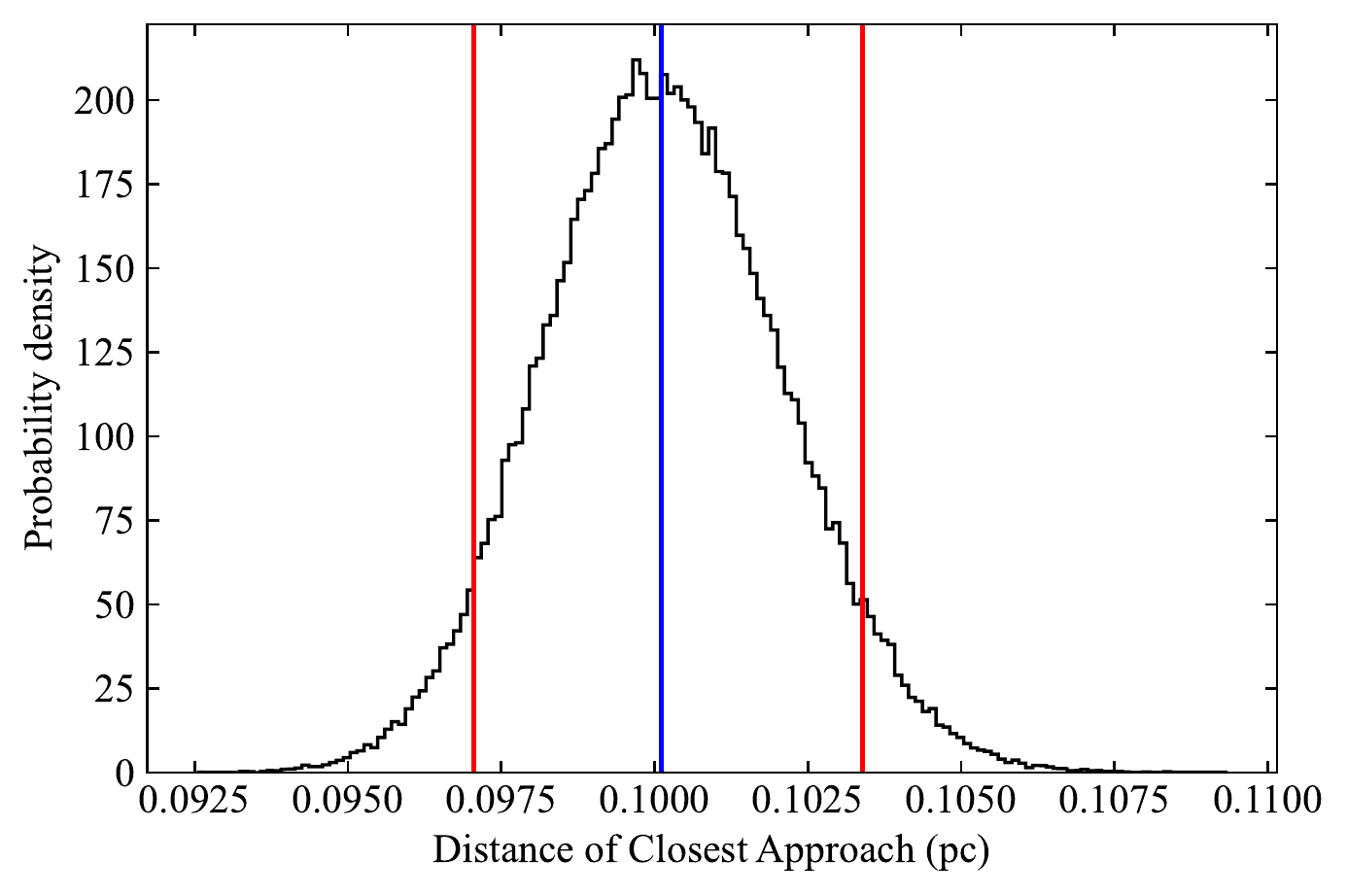}
            \includegraphics[scale=0.6,angle=0]{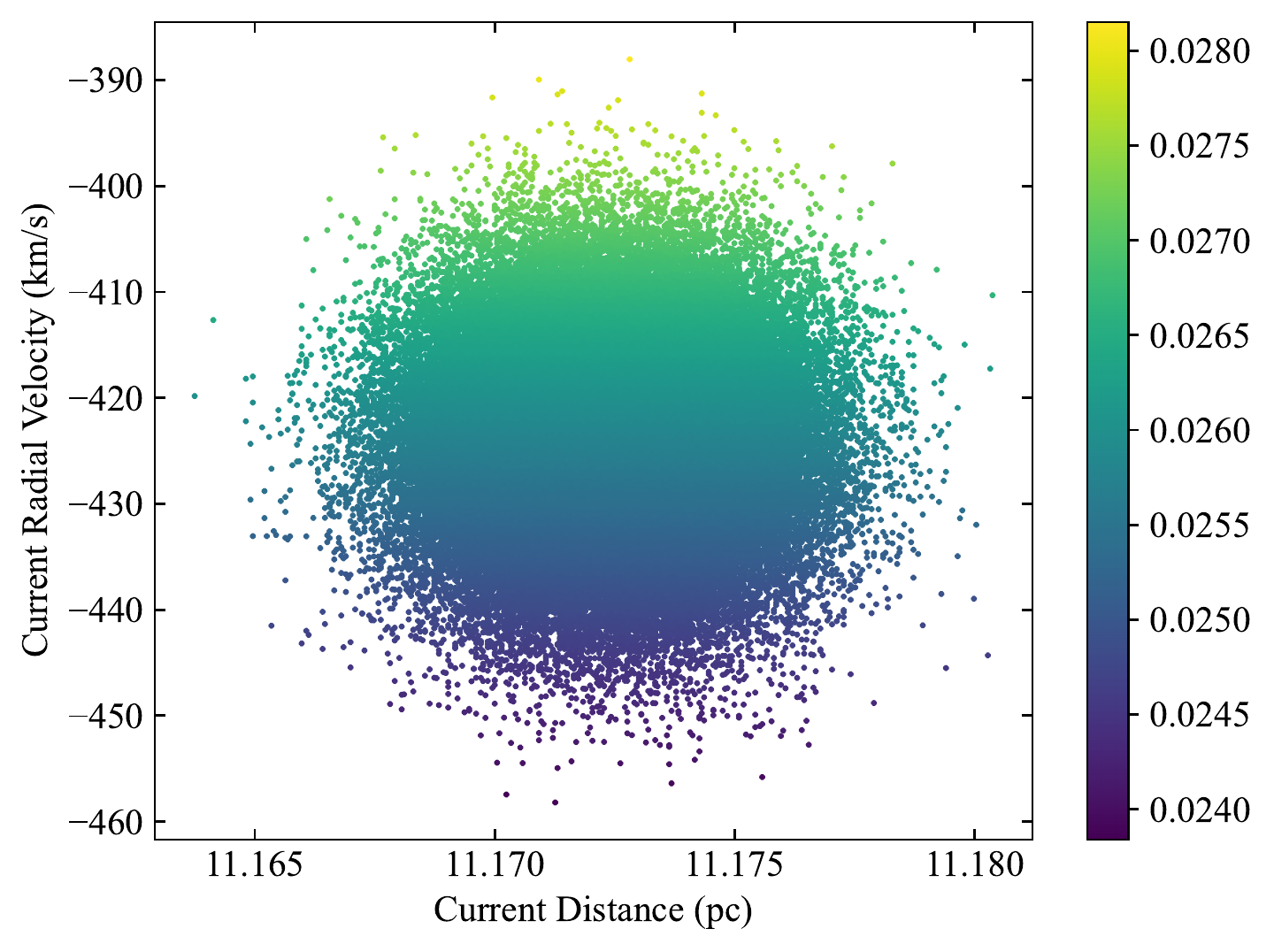}
            \includegraphics[scale=0.6,angle=0]{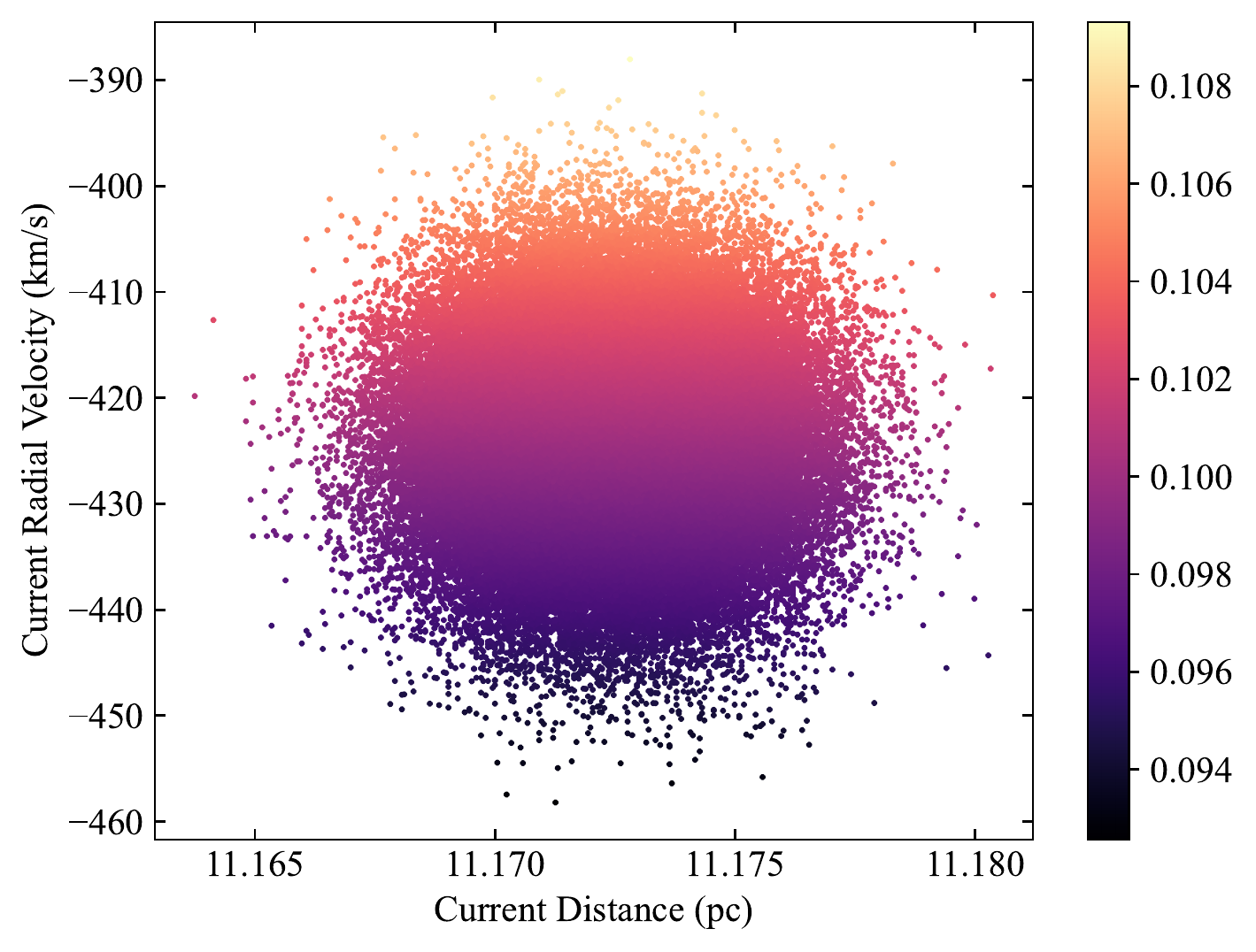}
            \caption{Same as Fig.~\ref{flyby} but considering the effect of Einstein's gravitational redshift \citep{1967ApJ...149..283G} on the value 
                     of the radial velocity (so it becomes $-423.74\pm8.18$~km~s$^{-1}$), as pointed out by \citet{2022arXiv220614443B}. The blue 
                     vertical lines mark the median values, and the red ones show the 5th and 95th percentiles.
            \label{flyby+}}
         \end{center}
      \end{figure*}

         Considering the effect of Einstein's gravitational redshift \citep{1967ApJ...149..283G} on the value of the radial velocity (so it becomes 
         $-423.74\pm8.18$~km~s$^{-1}$) as pointed out by \citet{2022arXiv220614443B} leads to similar values (see Fig.~\ref{flyby+}). In this case, 
         the distribution of distances of closest approach has a median value of 0.100~pc with a 90\% probability of coming within 0.097--0.103~pc of 
         the Sun; the associated time of perihelion passage is determined to be between 0.0250 and 0.0266~Myr with 90\% confidence, with a most likely 
         value of 0.0258~Myr.
%
%-----------------------------------------------------------------------------------------------------------------------------------------------------
%
%
%-----------------------------------------------------------------------------------------------------------------------------------------------------
%

      \subsection{Alternative scenarios}
\begin{figure*}
         \begin{center}
            \includegraphics[scale=0.6,angle=0]{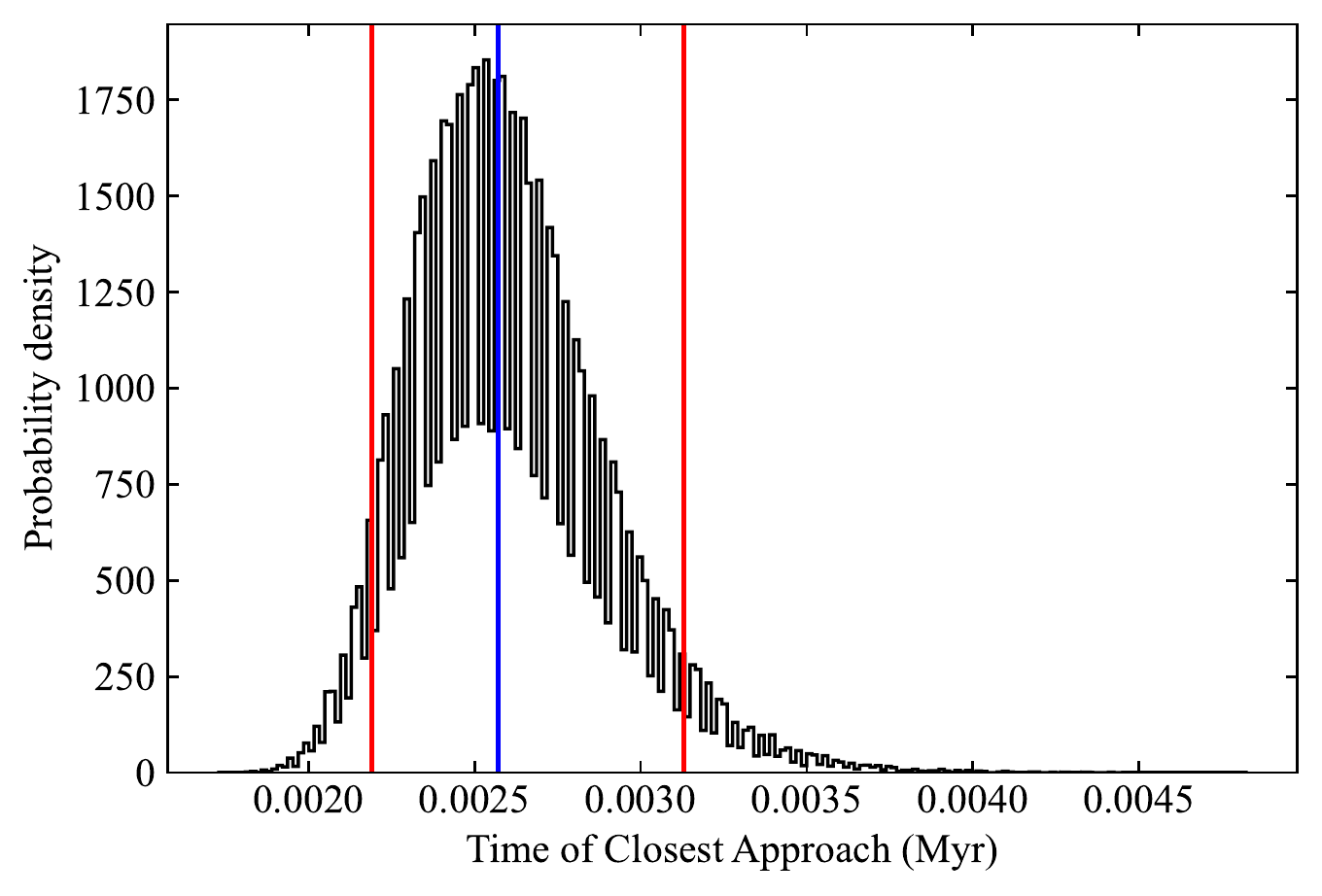}
            \includegraphics[scale=0.6,angle=0]{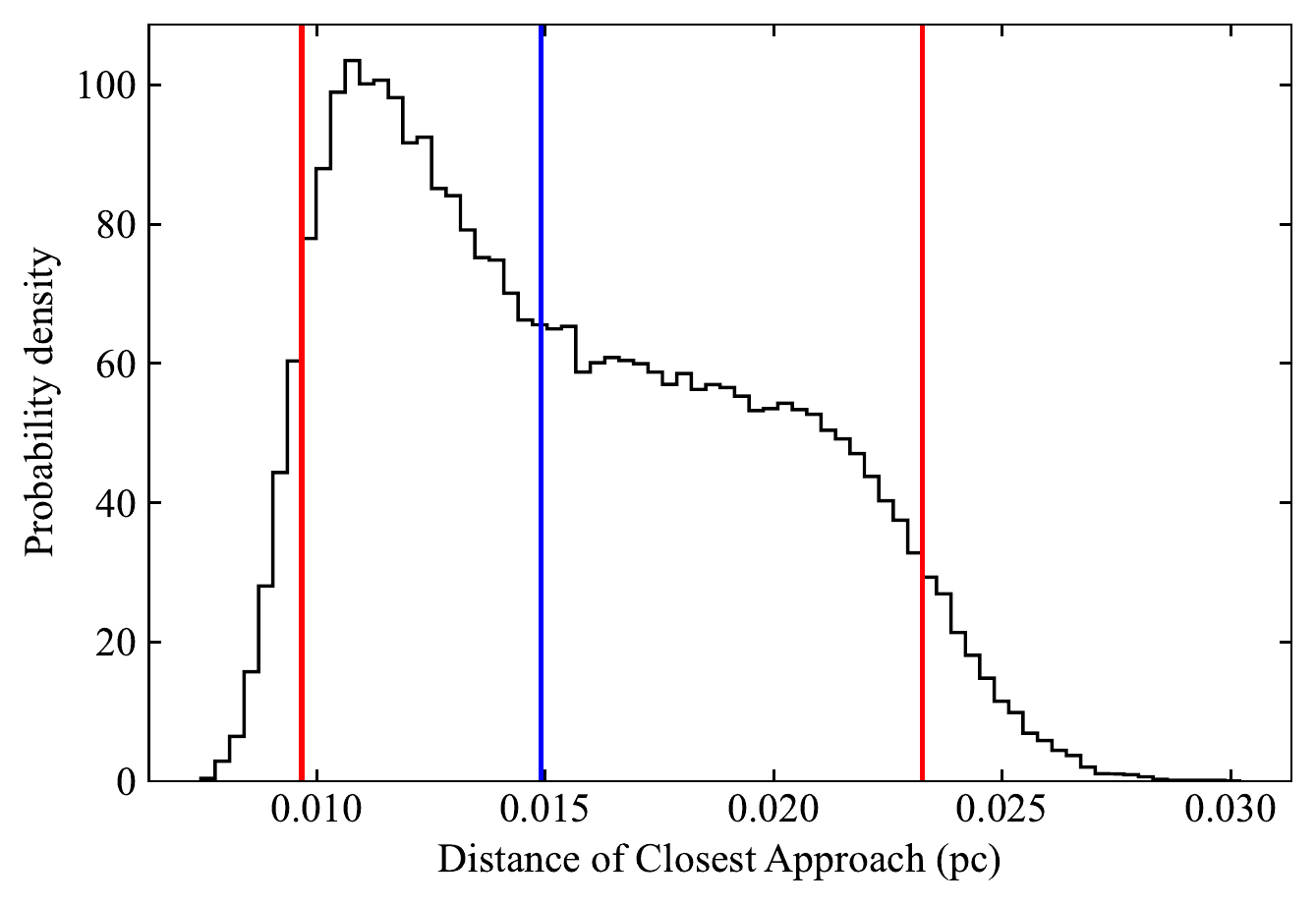}
            \includegraphics[scale=0.6,angle=0]{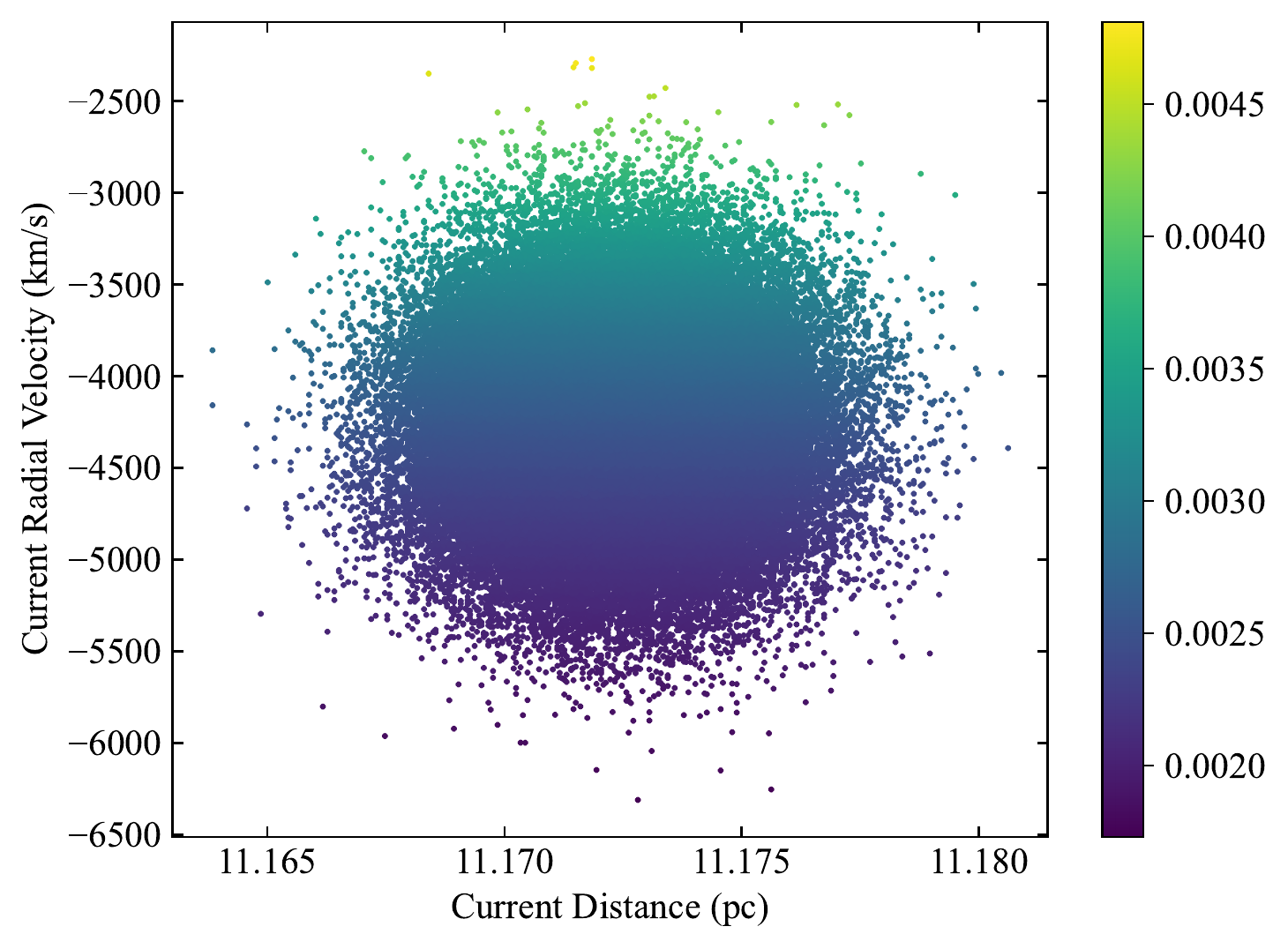}
            \includegraphics[scale=0.6,angle=0]{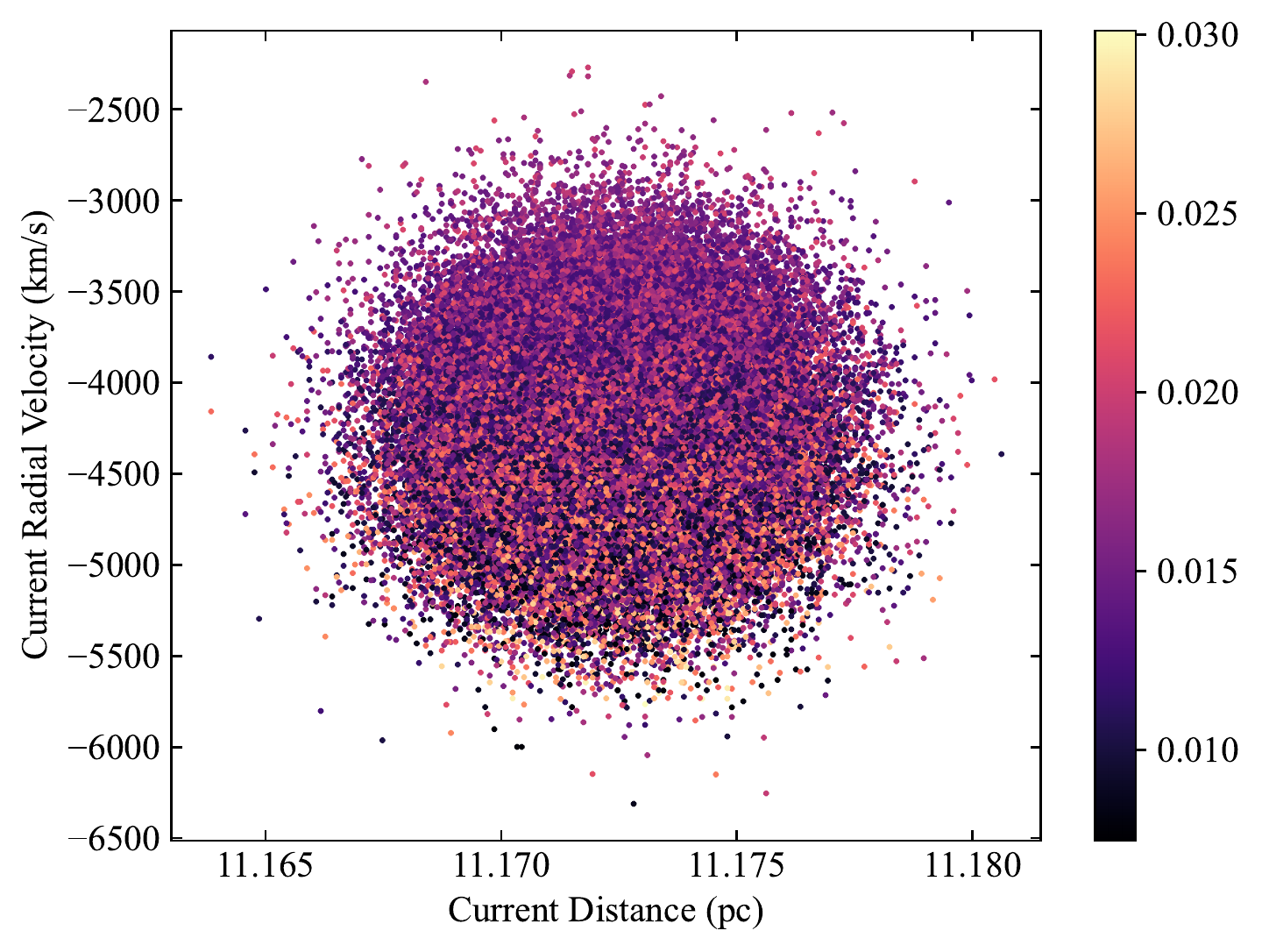}
            \caption{Same as Fig.~\ref{flyby} but assuming a radial velocity of $-4248\pm457$~km~s$^{-1}$, as discussed in 
                     Sect.~\ref{Controversy}. The blue vertical lines mark the median values, and the red ones show the 5th and 95th percentiles.
            \label{flyby++}}
         \end{center}
      \end{figure*}

\begin{figure*}
         \begin{center}
            \includegraphics[scale=0.6,angle=0]{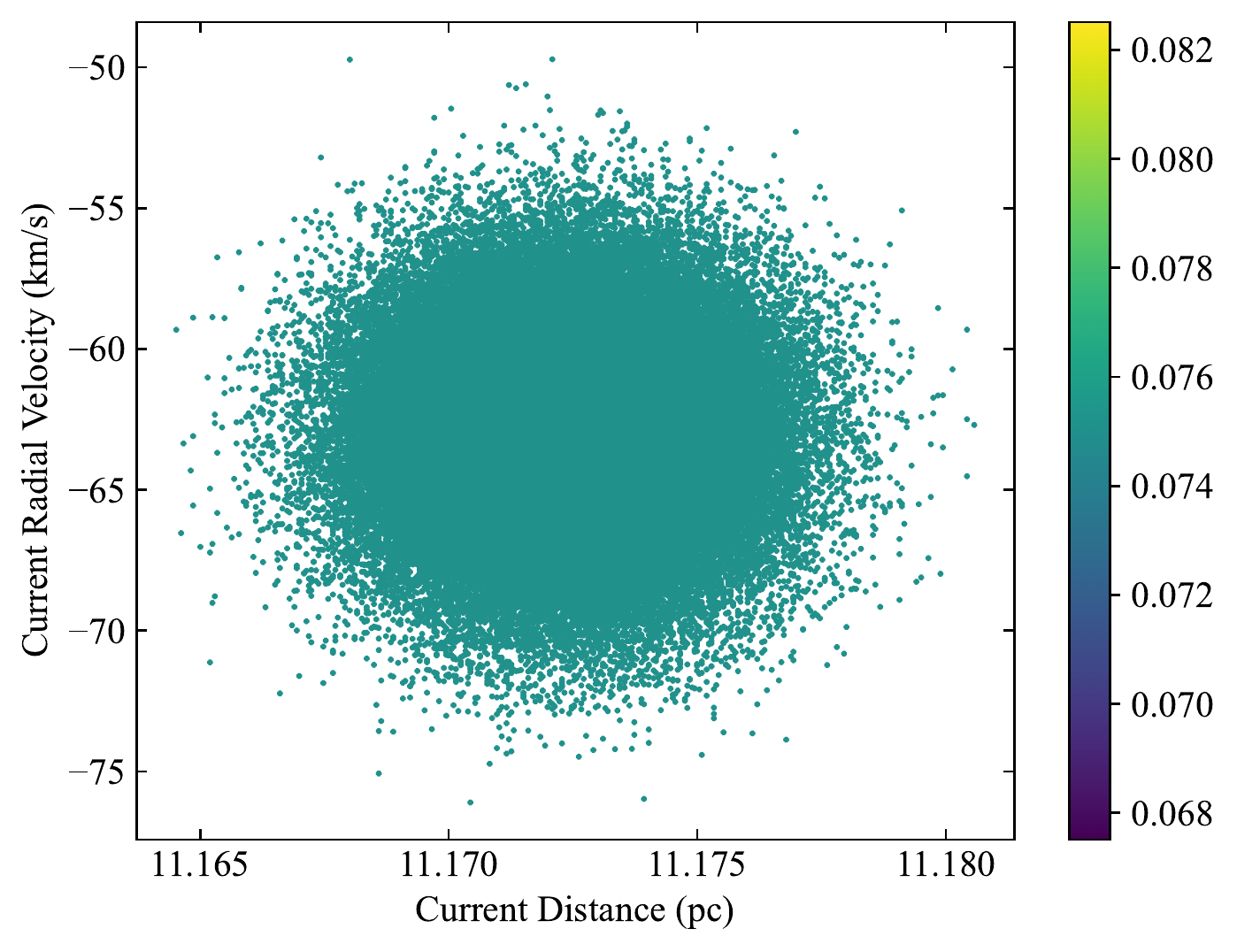}
            \includegraphics[scale=0.6,angle=0]{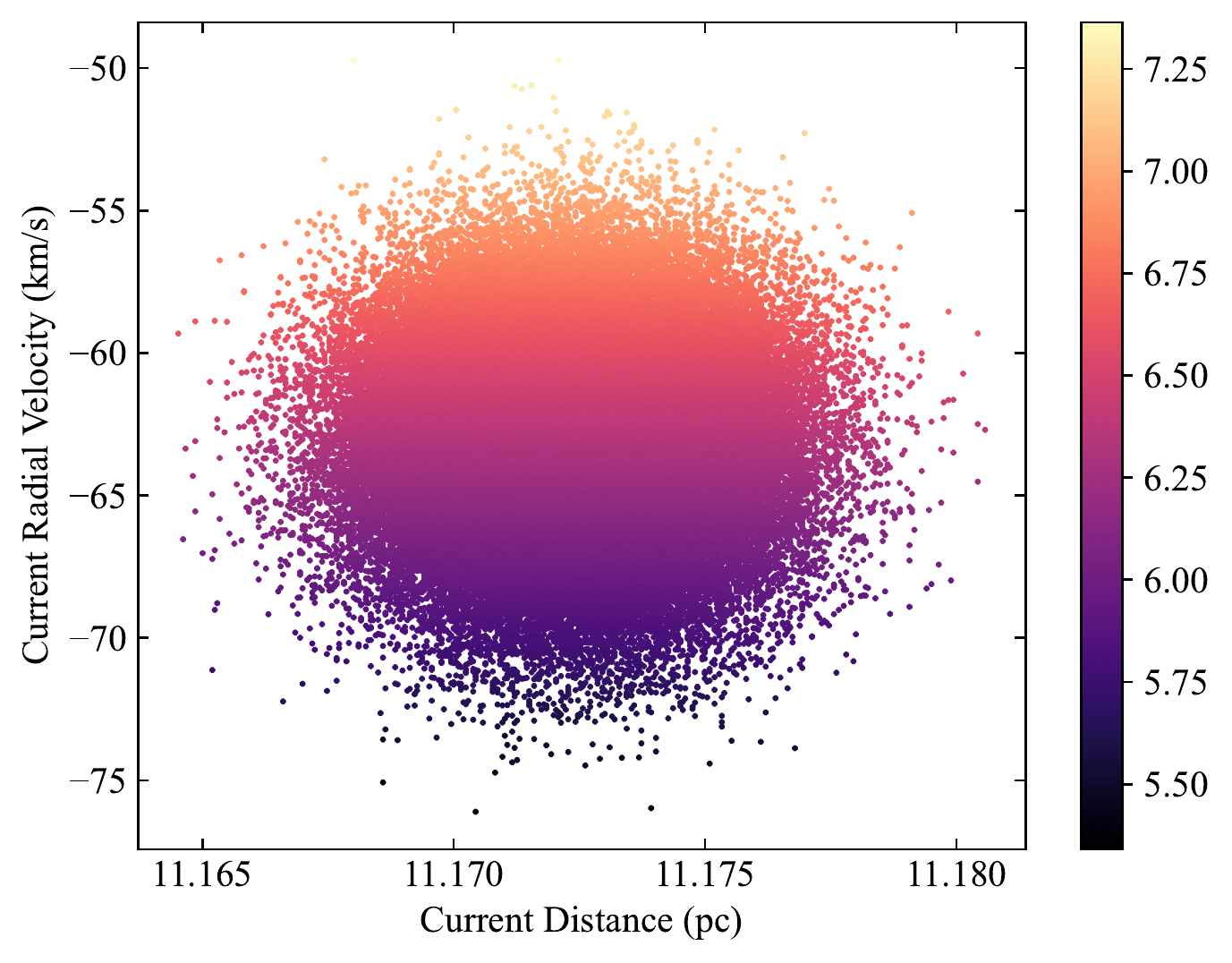}
            \caption{Same as the bottom panels of Fig.~\ref{flyby}, but assuming a radial velocity of $-63\pm3$~km~s$^{-1}$, as discussed in 
                     Sect.~\ref{Controversy}. 
            \label{flyby-}}
         \end{center}
      \end{figure*}

         Under the assumption that WD~0810-353 has a radial velocity of $-4248\pm457$~km~s$^{-1}$, as discussed in Sect.~\ref{Controversy}, 
         Fig.~\ref{flyby++} shows the results of 10$^{5}$ integrations of control orbits of WD~0810-353. In this case, the distribution of distances 
         of closest approach has a median value of 0.015~pc (or 3094~AU) with a 90\% probability of coming within 0.010--0.023~pc of the Sun; the 
         associated time of perihelion passage is determined to be between 0.0022 and 0.0031~Myr with 90\% confidence, with a most likely value of 
         0.0026~Myr. On the other hand, when a radial velocity of $-63\pm3$~km~s$^{-1}$ is considered, a distant encounter at 6.35~pc takes place in 
         0.075~Myr (see Fig.~\ref{flyby-}). As pointed out in the previous section, the faster the approach, the deeper the flyby.
%
%-----------------------------------------------------------------------------------------------------------------------------------------------------
%
%
%-----------------------------------------------------------------------------------------------------------------------------------------------------
%
%
%-----------------------------------------------------------------------------------------------------------------------------------------------------
%
%
%-----------------------------------------------------------------------------------------------------------------------------------------------------
%

   \section{Discussion\label{Discussion}} 
\begin{figure}
        \centering
         \includegraphics[width=\linewidth]{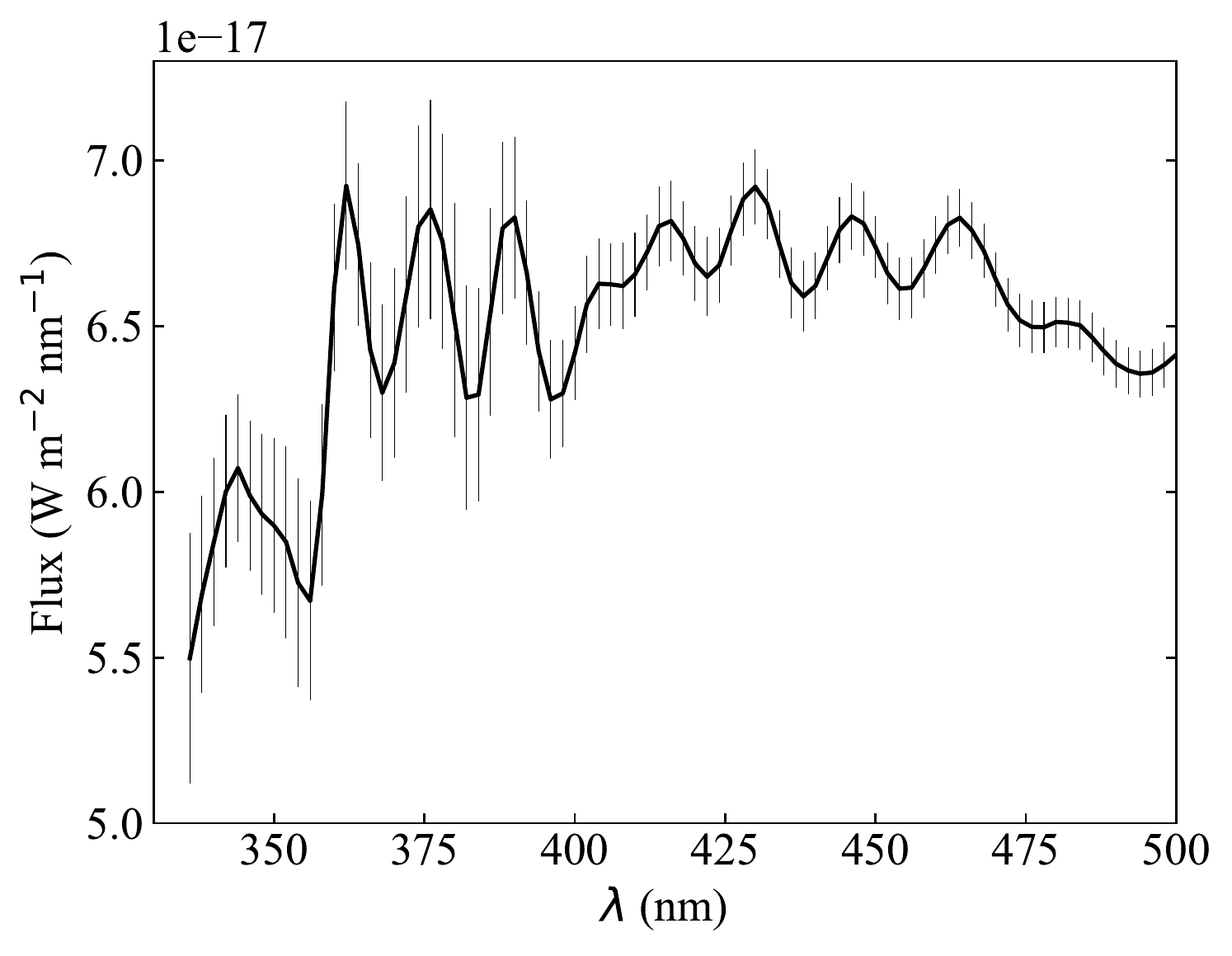}
         \caption{Detail of the low-resolution mean BP/RP spectrum of WD~0810-353. This part of the spectrum closely resembles that of LP~40-365 in 
                  Figs.~1 and 4 of \citet{2018ApJ...858....3R}.
                 }
         \label{spectrumWD0810-353}
      \end{figure}

      Using the data in \textit{Gaia}~DR3 as input, our analyses confirm that the magnetic white dwarf WD~0810-353 will experience a near-future close 
      encounter with the Solar System and that it is a probable member of a very exotic group of objects, the hypervelocity runaway white dwarfs. The 
      question then arises as to how we know if it is a survivor of a peculiar thermonuclear supernova. The prototype of this class of unusual objects 
      is LP~40-365, which has a rather peculiar spectral appearance (see Figs.~1 and 4 in \citealt{2018ApJ...858....3R}). A detailed plot of the XP 
      spectrum of WD~0810-353 (see Sect.~\ref{Controversy}) is shown in Fig.~\ref{spectrumWD0810-353}, and it closely resembles that of LP~40-365 in 
      Figs.~1 and 4 of \citet{2018ApJ...858....3R}. We therefore conclude that, if the radial velocity of WD~0810-353 in \textit{Gaia}~DR3 is not 
      spurious (but see Sect.~\ref{Controversy}), the nature (both physical and dynamical) of WD~0810-353 could be very similar to that of LP~40-365, 
      which is also unbound to the Galaxy.
%
%
%-----------------------------------------------------------------------------------------------------------------------------------------------------
%
%
%-----------------------------------------------------------------------------------------------------------------------------------------------------
%

      However, our discussion in Sect.~\ref{Controversy} and our initial comment in Sect.~\ref{Intro} both suggest that a flyby at a relative velocity 
      close to 70~km~s$^{-1}$ is far more likely, although it will lead to an unremarkable close encounter. The analyses in Sect.~\ref{Controversy} 
      also show the potential of \textit{Gaia}'s XP spectra to reproduce (albeit with lower quality) radial velocity results from the literature but 
      also to uncover candidates with unusual kinematics, particularly when \textit{Gaia}~Data Release 4 is released as it will feature a much more 
      extensive radial velocity catalog. In the absence of other data, the low-resolution XP spectra indeed appear to have the potential to allow the 
      preliminary identification of sources with unusual redshifts or blueshifts suitable for further study at higher spectral resolutions.  

   \section{Summary and conclusions\label{Conclusions}}
      In this paper we made use of input data from \textit{Gaia}~DR3 to study the evolution of the magnetic white dwarf WD~0810-353 forward in time 
      using direct $N$-body simulations and factoring the uncertainties into the calculations. In addition, we explored a tentative connection between 
      WD~0810-353 and the small group of known hypervelocity runaway white dwarfs using both astrometric and spectroscopic \textit{Gaia}~DR3 data. We
      also performed a detailed comparison between the values of the radial velocity of white dwarfs from the literature and from \textit{Gaia}~DR3, 
      providing a list of relevant caveats to be considered when dealing with \textit{Gaia}~DR3 radial velocity data. Our conclusions can be 
      summarized as follows.
      \begin{enumerate}
         \item In general, radial velocity determinations for white dwarfs in \textit{Gaia}~DR3 may be incorrect. This is due to the lack of synthetic 
               white dwarf templates in the RVS pipeline. However, we have found a number of examples in which the values from the literature are 
               statistically consistent with those from \textit{Gaia}~DR3, including the outlier LB~3209.
         \item We confirm that WD~0810-353 is single and not variable within the uncertainties associated with TESS data.
         \item We confirm that WD~0810-353 will experience a near-future Solar 
       System flyby with the following parameters (average and standard deviation):  a minimum approach distance of
               0.114$\pm$0.002~pc and a time of perihelion passage of 0.0292$\pm$0.0006~Myr. The relative velocity will be 
               high enough to preclude any significant perturbation on the Oort cloud. This conclusion is valid if the value of the radial velocity
               of this white dwarf in \textit{Gaia}~DR3 is not spurious.
         \item Considering the data at face value, we find that WD~0810-353 could be a hypervelocity runaway white dwarf as the value of its $V$ 
               component is above 600~km~s$^{-1}$, which is typical for ejected thermonuclear supernova survivors. 
         \item Again assuming a correct value of the radial velocity in \textit{Gaia}~DR3 and considering the value for the escape speed in the 
               neighborhood of the Sun, WD~0810-353 may already be unbound to the Milky Way. An origin in the Galactic disk is favored.
         \item The analysis of the low-resolution mean BP/RP spectrum of WD~0810-353 shows that it is a reasonable match for LP~40-365.
         \item If the radial velocity of WD~0810-353 in \textit{Gaia}~DR3 is spurious, our calculations show that an average radial velocity leads to
               an unremarkable flyby. If an absorption feature in the XP spectrum of this object is the H$\alpha$ line, then it may be an extreme
               hypervelocity white dwarf that will experience a deep and exceptionally fast flyby with the Solar 
       System a few thousand years from
               now.
      \end{enumerate}
      Given the potential importance of WD~0810-353, an independent radial velocity determination for this object will put an end to this controversy
      and improve our practical understanding of how reliable the radial velocity determinations for white dwarfs in \textit{Gaia}~DR3 are.

   \begin{acknowledgements}
      We thank three anonymous reviewers for input (one of them was particularly informative and the current form of this work owes a great deal to 
      the recommendations in this second review), S.~J. Aarseth for providing one of the codes used in this research, S. Deen for suggesting 
      UCAC4~398-010797 as a spectral analog for WD~0810-353, and A.~I. G\'omez de Castro for providing access to computing facilities. This work was 
      partially supported by the Spanish `Agencia Estatal de Investigaci\'on (Ministerio de Ciencia e Innovaci\'on)' under grant PID2020-116726RB-I00 
      /AEI/10.13039/501100011033. In preparation of this paper, we made use of the NASA Astrophysics Data System and the ASTRO-PH e-print server. This 
      research has made use of the SIMBAD database, operated at CDS, Strasbourg, France. This work has made use of data from the European Space Agency 
      (ESA) mission \textit{Gaia} (\url{https://www.cosmos.esa.int/gaia}), processed by the \textit{Gaia} Data Processing and Analysis Consortium 
      (DPAC, \url{https://www.cosmos.esa.int/web/gaia/dpac/consortium}). Funding for the DPAC has been provided by national institutions, in 
      particular the institutions participating in the \textit{Gaia} Multilateral Agreement.
   \end{acknowledgements}

   \bibliographystyle{aa}

\end{document}